\documentclass[aps,prl,twocolumn]{revtex4-1}

\usepackage{graphicx}
\usepackage{dcolumn}
\usepackage{bm}

\usepackage{subcaption}
\usepackage{amsfonts}
\usepackage{amsmath}
\usepackage{color}
\usepackage{amssymb}
\usepackage{tikz}

\usepackage{algorithm}
\usepackage[noend]{algpseudocode}
\usepackage{braket}
\usepackage[export]{adjustbox}
\usepackage{footnote}
\usepackage{multirow}

\begin{document}


\title{Bayesian optimization for tuning and selecting hybrid-density functionals}

\author{R. A. Vargas--Hern\'andez}
 \email{ravh011@chem.ubc.ca}
\affiliation{Department of Chemistry, University of British Columbia, Vancouver, British Columbia, Canada, V6T 1Z1.}
\altaffiliation[Current Address: ]{Chemical Physics Theory Group, Department of Chemistry, University of Toronto, Toronto, ON Canada M5S 3H6.}%


\date{\today}

\begin{abstract}
The accuracy of some density functional (DF) models, widely used in material science, depends on empirical or free parameters which are commonly tuned using reference physical properties.
The optimal value of the free parameters is regularly found using grid search algorithms, which computational complexity scales with the number of points in the grid.
In this report, we illustrate that Bayesian optimization (BO), a sample-efficient machine learning algorithm,  can efficiently calibrate different density functional models, e.g., hybrid-exchange-correlation and range-separated density functionals. 
We present that, BO can optimize the free parameters of hybrid-exchange-correlation functionals, with approximately 55 evaluations of the root-mean-square or mean-absolute error functions of the atomization energies and the bond length of the Gaussian-1 (G1) database. 
We also illustrate that BO can identify, without any prior information, the most appropriate exchange-correlation functional by navigating through the space of density functional models. 
We optimize and select the free parameters and the exchange-correlation functional form jointly by also minimizing the root-mean-square error function with respect to the atomization energies of the G1 database using BO.
\end{abstract}

\pacs{Valid PACS appear here}
\maketitle

\emph{Introduction.} Computational models based on density functional theory (DFT) is the workhorse of quantum mechanical simulations for predicting structures, energetics, and other physical properties across different fields.
While DFT is in principle an exact theory, most of density functional (DF) models are not considered \emph{ab initio} methods as they contain empirical parameters \cite{dftbook}. 
Over the last decades a great variety of DF models have been proposed \cite{dft_revmodphys,dft_HGordon_benchmark}.
With the large variety of DF models it is critical to understand which model better predicts the physical properties of the system of interest. 
This has lead to a large number of benchmark studies where different DF models are compared \cite{dft_HGordon_benchmark,DFT_benchmark0,DFT_benchmark1,DFT_benchmark2}. 
However, due to the wide range of applications of DF models in material science, an automated model selection method is necessary for an accurate prediction of molecular properties \cite{DFT_to_ML}.

The selection of physical models can be formulated as an optimization problem where the free parameters of a model, denoted as $\mathbf{x}$, are tuned to best reproduce the physical properties \cite{ML_optbook,ML_opt}. 
Commonly a loss or cost function, ${\cal L}$, is used to determine the relation between the free parameters and the accuracy of the models. 
The minimizer of ${\cal L}$ is the value of the free parameters for the most accurate model,
\begin{eqnarray}
\mathbf{x^*} = \underset{\mathbf{x}\in \mathbb{R}^\ell}{\mathrm{arg\;min}}  \;{\cal L}(\mathbf{x}). 
\label{eqn:min_f}
\end{eqnarray}
Any ${\cal L}$ function can be minimized using gradient-based methods \cite{ML_opt,ML_optbook} by computing the change of ${\cal L}$ with respect to its parameters. 
For DF models, the gradient of ${\cal L}$ with respect to any of the free parameters $x_i$ may not have a simple analytical form since it also depends on the physical property, denoted as $R_{m_i}$, chosen to evaluate the accuracy of the model, $\frac{\partial {\cal L} }{\partial x_i} \propto \frac{\partial R_{m_i} }{\partial x_i}$.
Because of the complexity to compute the gradient of ${\cal L}$ with respect the $x_i$, grid search methods are the common approach to optimize DF \cite{dftbook,B_acm3_0,B_acm3_1,OptB3LYP}. 
In this letter we present a scheme based on a machine learning (ML) algorithm to efficiently screen different DF models.  

ML algorithms have been demonstrated to be powerful numerical tools to simulate many-body physics \cite{vonLilienfeld,vonLilienfeld_2,Aspuru_NNmolecules,Wang_prb,Carleo_science,Carrasquilla_natphys,ravh_prl,Aspuru_NNmolecules_2}, e.g., reducing the computational complexity in DF calculations by bypassing the Kohn-Sham equations \cite{KE_DFT,ML_DFT,ML_DFT_2}.
Likewise, Bayesian ML models have been used to study quantum systems \cite{RK_bayesML}; for instance, the optimization for producing Bose-Einstein condensates \cite{GPBEC}, simulate chemical reactions \cite{Reiher_chemrxn_error,Reiher_chemrxn_error2} and controlling a robot to do chemical synthesis \cite{phoenics,chemos}, both using a probabilistic regression model.
Additionally, the error of DF models has also been estimated using Bayesian statistics \cite{BEE_PRL,BEE_PRB,Bayes_DFT_0,Bayes_DFT_1,Walter_BEE,mBEEF,Bayes_DFT_water}.

One of the most important results of ML are the optimization methods designed to minimize complex functions to train and select ML models \cite{ML_optbook,ML_opt}.
Bayesian optimization (BO) is one of the most common ML algorithms used to minimize functions whose gradients can not be computed \cite{BO_Adams,BO_Freitas}. 
In the field of molecular physics, BO was recently used to generate low-energy molecular conformers \cite{BO_PES,BO_geo}, and to build global potential energy surfaces for reactive molecular systems using feedback from quantum scattering calculations \cite{ravh_njp}.
BO has also been applied to efficiently screen for chemical compounds \cite{BO_mat,BO_dft_calc,BO_phonono_transport}, and to minimize the energy function for the Ising model \cite{BO_ising}.
In this contribution we demonstrate that the parameters of a DF model can be efficiently tuned using BO. 
We consider two cases, i) the search for the optimal values of the free parameters DF, e.g., hybrid exchange-correlation (XC) and range-separated functionals, and ii) the search for the most accurate XC functional form. 
The value of the methodology presented here is clear since automation to improve DF models is one of the most important goals in computational physics, chemistry and material science.\\

\emph{Method.}
Bayesian optimization is a sequential search algorithm designed to find the global minimizer (or maximizer) of an unknown non-analytic or oracle function, Eq. \ref{eqn:min_f}.
BO requires two components: a model that approximates ${\cal L}$ and an acquisition function, $\alpha(\mathbf{x})$ \cite{BO_Adams,BO_Freitas}.
Here we use Gaussian process (GP) models as the probabilistic model to approximate ${\cal L}$ \cite{gpbook}.
GP model is a non-parametric regression model $f(\mathbf{x})$, whose function values $f(\mathbf{x}_1),\cdots,f(\mathbf{x}_N)$ are jointly Gaussian distributed.
The prediction of a new point using GP models is carried out by computing the conditional distribution of ${\cal L}(\mathbf{x_*})$ given training data, denoted as ${\cal D} = \{X,\mathbf{y}\}$. 
The conditional distribution has a closed form characterized by its mean, $\mu(\mathbf{x_*})$, and standard deviation, $\sigma(\mathbf{x_*})$,
\begin{eqnarray}
\mu(\mathbf{x^*}) &=& K(\mathbf{x_*},X)^\top \left [K(X,X) + \sigma_n^2 I\right ]^{-1}\mathbf{y} \label{eq:mean_gp}\\
\sigma(\mathbf{x^*}) &=& K(\mathbf{x_*},\mathbf{x_*}) \nonumber \\  &&- K(\mathbf{x_*},X)^\top\left [K(X,X) + \sigma_n^2 I\right ]^{-1}K(\mathbf{x_*},X), \label{eq:std_gp} 
\end{eqnarray}
where $K(\cdot,\cdot)$ is the design or covariance matrix with  matrix elements $K_{i,j} = k(\mathbf{x}_i,\mathbf{x}_j)$, where $k(\cdot,\cdot)$ is the kernel function. For this work we used the \emph{radial basis function} (RBF) kernel,
 \begin{eqnarray}
k_{RBF}(\mathbf{x}_i,\mathbf{x}_j) = \exp\left(-\frac{1}{2}(\mathbf{x}_i-\mathbf{x}_j)^\top M (\mathbf{x}_i-\mathbf{x}_j)\right),
\label{eqn:rbf_kernel}
\end{eqnarray}
where $M$ is a diagonal matrix that has different length-scale parameter, $\ell_d$, for each dimension of $\mathbf{x}$. All $\ell_d$ are described as $\boldsymbol{\theta}$. 
The parameters of the kernel function are optimized by maximizing the log marginal likelihood,
 \begin{eqnarray}
\log p(\mathbf{y}|X,\mathbf{\theta}) &=& -\frac{1}{2}\mathbf{y}^\top K(X,X)^{-1} \mathbf{y} \nonumber\\
&& -\frac{1}{2}\log|K(X,X)| - \frac{N}{2}\log(2\pi),
\end{eqnarray}
where $N$ is the total number of points in ${\cal D}$ and $|K(X,X)|$ is the determinant of the design matrix. 
For more details on GP models, see Refs. \cite{gpbook,sm_gp}.

The goal of BO is to reduce the computational complexity of minimizing ${\cal L}$, by iteratively minimizing $\alpha(\mathbf{x})$, which is less computationally demanding \cite{BO_Adams,BO_Freitas}.
In BO, the acquisition function quantifies the informational gain if ${\cal L}$ were evaluated at a new point, ${\cal L}(\mathbf{x}_{N+1})$. 
Here, we only considered two acquisition functions: the \emph{expected improvement} (EI), 
\begin{eqnarray}
\alpha_{EI}(\mathbf{x}) &=&( \mu(\mathbf{x}) - y_{max})\Phi(z(\mathbf{x};y_{max})) + \sigma(\mathbf{x})\phi(z(\mathbf{x};y_{max})), 
\label{eqn:a_ei}\nonumber \\
\end{eqnarray}
where $z(\mathbf{x};y_{min}) = (\mu(\mathbf{x})-y_{min})/ \sigma(\mathbf{x})$, $\Phi(\cdot)$  is the normal cumulative distribution and $\phi(\cdot)$ is the normal probability distribution. $y_{min}$ is the minimum value observed in the training data, $y_{min} = \text{arg\;min }\mathbf{y}$.
Secondly, we considered  the \emph{upper confidence bound} (UCB),
 \begin{eqnarray}
\alpha_{UCB}(\mathbf{x}) &=& \mu(\mathbf{x})+ \kappa \sigma(\mathbf{x}), \label{eqn:a_ucb}
\end{eqnarray}
where $\kappa$ is the exploration-exploitation constant. For all the results presented in this work we set $\kappa = 1$.
For both acquisition functions, $\mu(\mathbf{x})$ and $\sigma(\mathbf{x})$ are the mean and the standard deviation from a GP model, Eqs. (\ref{eq:mean_gp}) and (\ref{eq:std_gp}). 
By sequentially minimizing the acquisition function and evaluating ${\cal L}$ in the proposed points, BO finds the minimum/maximum of a non-analytic function, such as  ${\cal L}$ \cite{SM_PBE0}. 

The most common DF methods are the hybrid-density functionals, introduced by Becke \cite{B_acm3_0,B_acm3_1}, which combine local and non-local treatments of \emph{exchange} (X) and \emph{correlation} (C) with the Hartree-Fock (HF) exchange,
\begin{eqnarray}
E^{acm3}_{XC} &=& E^{LSD}_{XC} + a_0(E^{exact}_{X} - E^{LSD}_{X}) + a_X(E^{GGA}_{X} - E^{LSD}_{X}) \nonumber \\
&&+ a_C(E^{GGA}_{C} - E^{LSD}_{C}),
\label{eq:acm3}
\end{eqnarray}
where $a_0$, $a_X$, and $a_C$ are adjustable parameters,  $E^{GGA}_{X}$ and $E^{GGA}_{C}$ are the generalized gradient approximation (GGA) exchange and correlation functionals, and $E^{LSD}$ is the local spin density (LSD) part. 
Hybrid functionals of the form of Eq. \ref{eq:acm3} are usually referred to collectively as ACM3 \cite{B_acm3_0,B_acm3_1,acm_theory}.
In the following section we present how BO can optimize the values of $a_0$, $a_X$, and $a_C$, and select the most accurate par of XC functional given a benchmark set of physical properties, $R^{exact}_{m_i}$.\\

\emph{Results.} 
The loss function we considered for the optimization of all the different DF models is the \emph{root mean square error} (RMSE) function,
\begin{eqnarray}
{\cal L}_{RMSE} &=& \sqrt{\frac{1}{|M|}\sum^{|M|}_{m_i} \left(\hat{R}_{m_i}(\mathbf{x}; {\cal M}_i) - R^{exact}_{m_i} \right )^2}, 
\label{eqn:loss}
\end{eqnarray}
where $R^{exact}_{m_i}$ is the atomization energy $m_i$ of the Gaussian-1 (G1) database \cite{G1,G1_1,G1_2} and $\hat{R}_{m_i}$ is the atomization energy predicted with a DF model ${\cal M}_i$. 
$\mathbf{x}$ are the free parameters of ${\cal M}_i$ and $|M|$ is the total number of physical properties used in the error function, $|M| = 32$.  
All the DF calculations are performed with the Gaussian 09 suite \cite{G09}, 
and the molecular geometries used in the DF calculations were optimized with MP2/6-31G($d$).

First we optimized the free parameter $a_0$ of PBE0 where, $a_X = 1 - a_0$ and $a_C = 1$ \cite{acm_theory0,acm_theory,PBE0_1} using BO with the UCB acquisition function with $\kappa =1$, and the basis set 6-31G($d$) .
With only 6 total evaluations of ${\cal L}$, BO found that the lowest RMSE is when $a_0 = 0.1502$; $\text{RMSE}=10.08$ kcal mol$^{-1}$. 
The value of the RMSE predicted with the original PBE0 \cite{acm_theory,PBE0_1}, $a_0 = \frac{1}{4}$, is  $11.30$ kcal mol$^{-1}$ \cite{SM_PBE0}.

We also considered the jointly optimization of the $a_0$, $a_X$, and $a_C$ for 30 different XC functionals using BO; combination of  5 different X functionals \cite{exchange}, $E_X$, and 6 different C functionals \cite{correlation}. 

\begin{figure}
  \centering
	\includegraphics[width=0.475\textwidth]{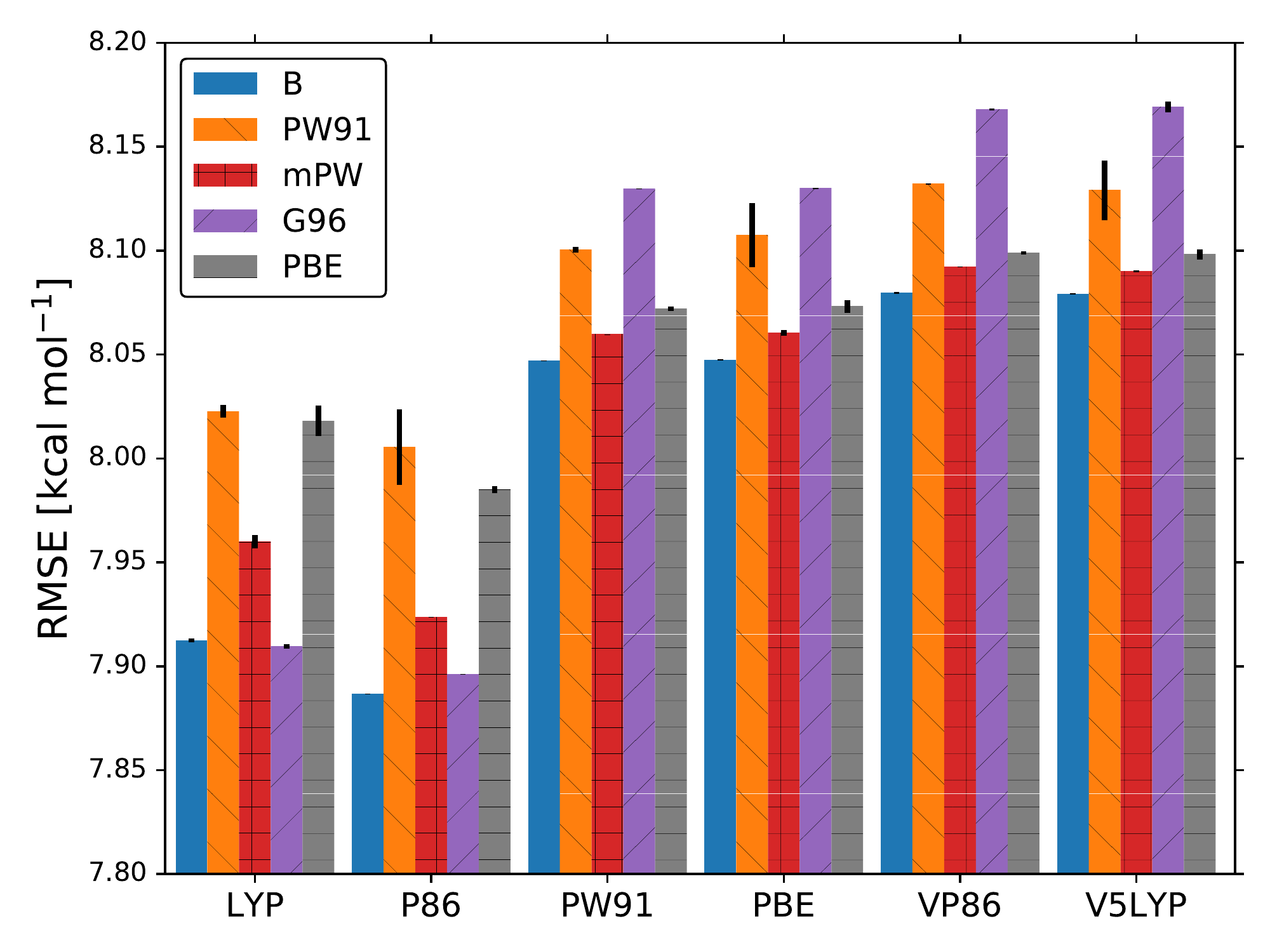}
   \caption{Lowest RMSE, Eq. \ref{eqn:loss}, found by BO averaged over 5 different optimizations for 30 different XC functionals where $R_{m_i}$ are atomization energies of the G1 molecules \cite{G1,G1_1,G1_2}.
	We considered 5 different exchange functionals, each X functional is color and hatch coded in the inset of the graph, and the horizontal axis labels represent the 6 different C functionals. 
    	For each XC functional, we optimized the free parameters of Eq. (\ref{eq:acm3}) by initializing BO with 5 different set of points, $N_0 = 15$, sampled with LHS. We use the UCB acquisition function with $\kappa = 0.1$  \cite{SM_acm3_tab}. The maximum number of points allowed in the BO algorithm was 70, including the initial points. The values of $a_0,a_X$ and $a_C$ for all 30 XC functionals are reported in Table III SM.
  }\label{fig:min_rmse_DFTall}
\end{figure}

For each XC functional we carried out 5 different optimizations with different 15 initial points, $N_0 = 5 \times d$  where $d$ is the dimensionality of ${\cal L}$, $d=3$.
These points were sampled using the latin hyper cube sampling (LHS) algorithm \cite{LHS} to avoid sampling multiple points close to each other. 
For all calculations we used 6-31G($d$) and the molecular geometries were optimized with MP2/6-31G($d$).
The lowest RMSE found by BO for each XC functional is displayed in Fig. \ref{fig:min_rmse_DFTall}. 
For each optimization, the maximum number of iterations allowed was 70 total points including the LHS points. 
The optimized coefficients for all 30 XC functionals are reported in Table III SM.

\begin{figure}
    \centering
        \includegraphics[width=0.45\textwidth]{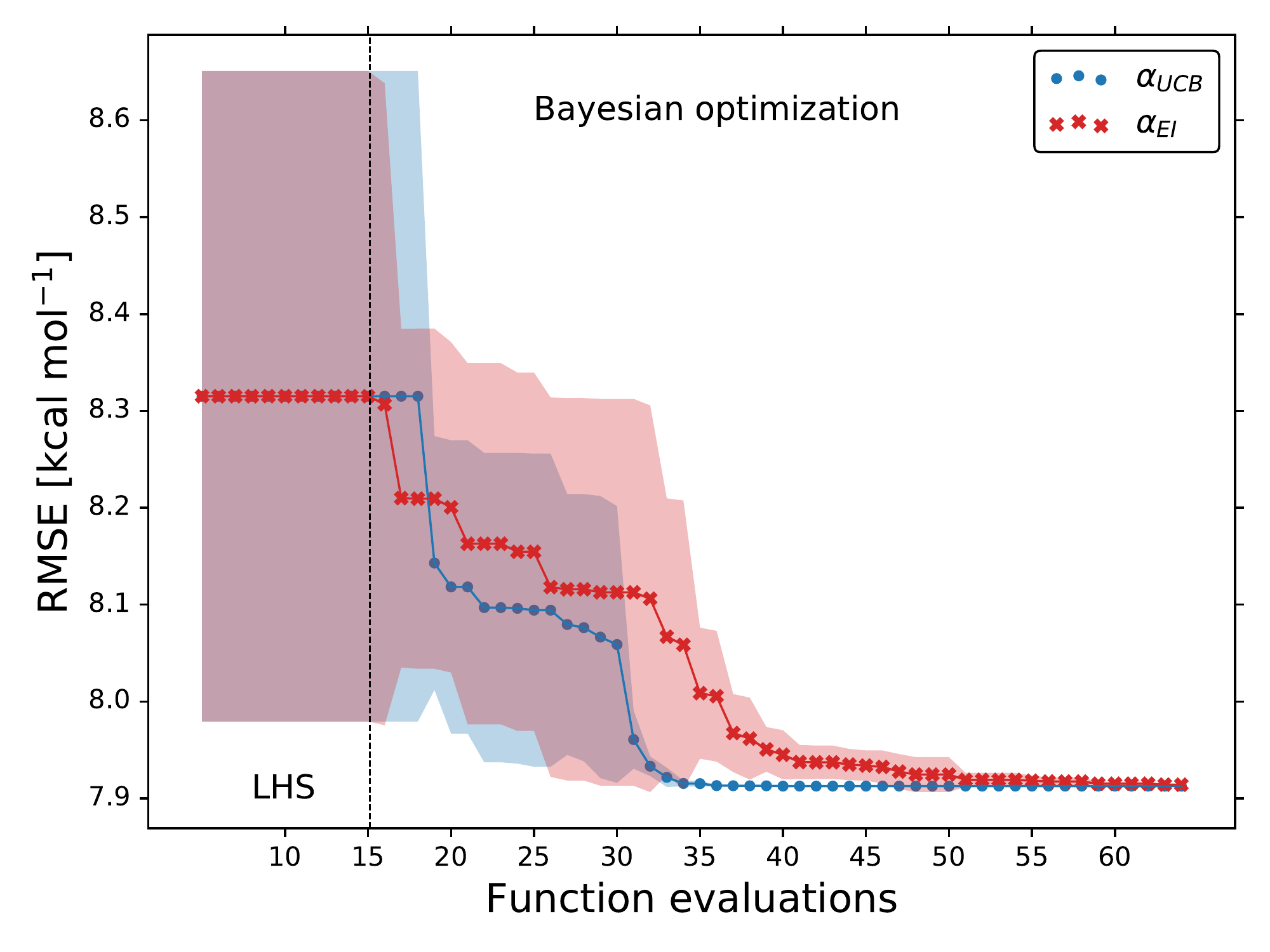}
        \includegraphics[width=0.45\textwidth]{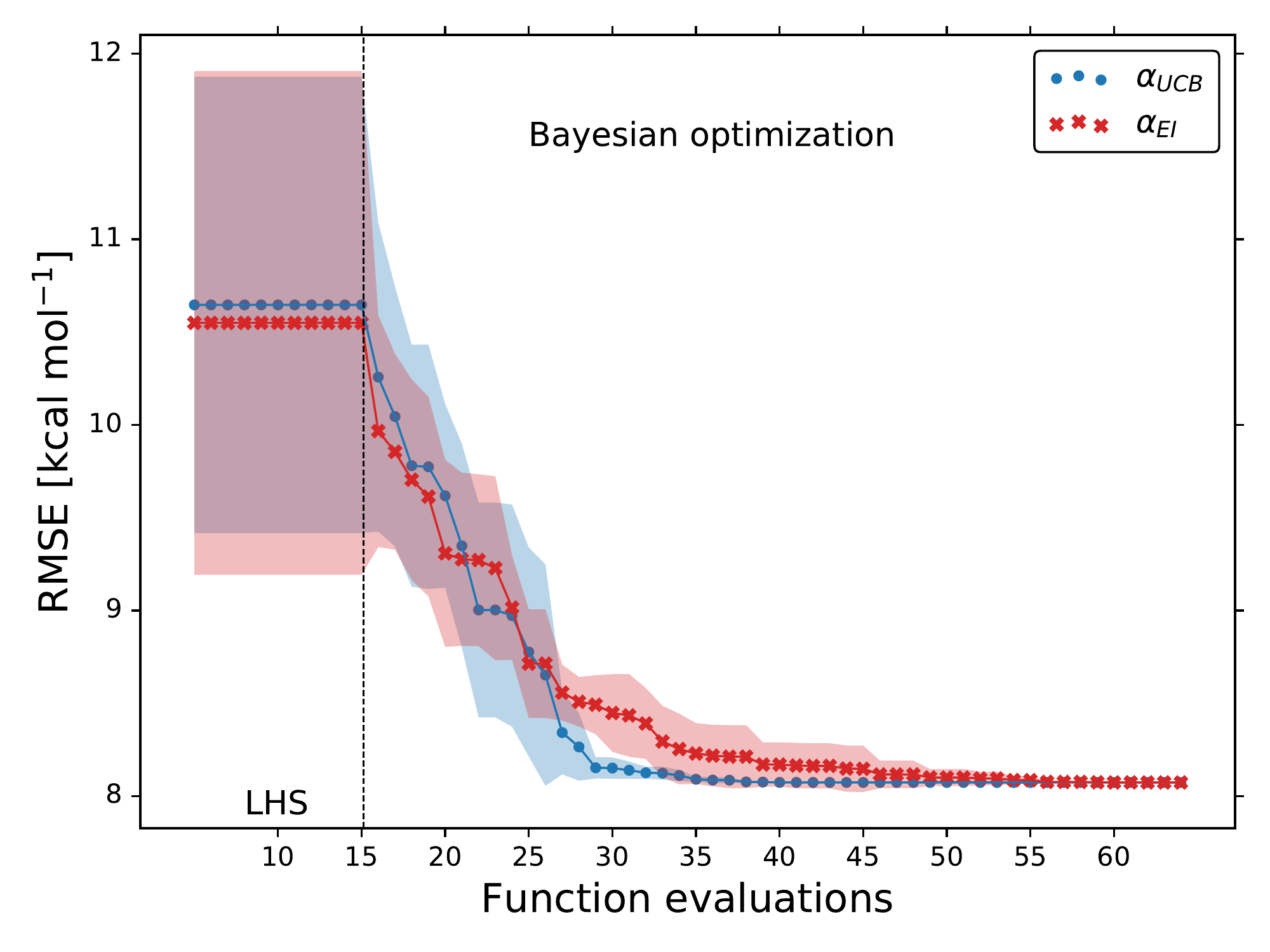}
    \caption{Lowest RMSE at each iteration of the BO algorithm for the optimization of the free parameters of Eq. (\ref{eq:acm3}) for the B-LYP (upper panel) and PBE-PBE (lower panel) functionals. We used two different acquisition functions: {\color{blue} $\bullet$}-symbol for the $\alpha_{UCB}$ with $\kappa=0.1$ and {\color{red} $\times$}-symbol for the $\alpha_{EI}$. 
           The RMSE is computed using Eq. \ref{eqn:loss}, where $R_{m_i}$ are atomization energies of the G1 molecules \cite{G1,G1_1,G1_2}.
	In each iteration of the BO algorithm, we plot the mean (symbols) and standard deviation (shaded area) for the lowest RMSE, averaged over 5 independent runs.
         For each optimization we considered 15 different initial points, sampled with LHS.}
\label{fig:bo_acq_acm3}
\end{figure}

As it is mentioned above, the goal of BO is to circumvent the optimization problem of ${\cal L}$ to a sequential optimization of $\alpha(\mathbf{x})$, which is a less computational demanding task.
In the case of the UCB function, the exploration-exploitation constant $\kappa$ allows us to probe the space of ${\cal L}$ without being trapped in a possible local minimum.
In the limit where $\sigma(\mathbf{x}) > \mu(\mathbf{x})$, the maximum of $\alpha_{UCB}$ is where the GP model is less certain, allowing us to explore the space. 
When $ \sigma(\mathbf{x}) <  \mu(\mathbf{x})$, $\alpha_{UCB}$ allows to explode and converge towards the minimum of ${\cal L}$ in the case for DF methods. 
The quality of points that are proposed by the acquisition function is a key component in BO \cite{BO_Adams, BO_Freitas}.
For instance, as shown in Fig. \ref{fig:bo_acq_acm3} we illustrate that as the number of iterations increase in the BO algorithm the GP model becomes more certain about where the minimum of ${\cal L}$ is located.
With approximately 50 total evaluations of ${\cal L}$, including the LHS points, BO found the optimal values of $a_0$, $a_X$, and $a_C$ for a given XC functional. 

We compare the efficiency of BO by using a grid with 725 points to search for the minimum of $a_0$, $a_X$, and $a_C$ for the B-LYP functional.  The lowest RMSE found was 8.17 kcal mol$^{-1}$. $\Delta a_i = 0.1$ is the space between points in the grid.
The RMSE for the B-LYP functional found by BO is 7.91 kcal mol$^{-1}$, and 8.07 kcal mol$^{-1}$ for PBE-PBE; both results were averaged over 5 different BO optimizations, Fig.  \ref{fig:bo_acq_acm3}.
$a_0 =  0.1100$, $a_X = 0.8293$ and $a_C = 0.0201$ are the optimized coefficients for the B-LYP functional; while for PBE-PBE, $a_0 = 0.1028$, $a_X = 0.8708$ and $a_C = 0.0292$, all obtained using BO, Table \ref{tab:a0_ax_ac_coeff}.
The RMSE for the functionals with their well known versions, B3LYP \cite{B3LYP, B3LYP_2} and PBE0 \cite{B_acm3_0}, are 9.48 kcal mol$^{-1}$ and 11.3 kcal mol$^{-1}$ respectively.  
It is important to note that the result of BO is independent of the initial set of points sampled with LHS, Fig. \ref{fig:bo_acq_acm3}. 
In Ref. \cite{OptB3LYP} the optimization of free parameters of B-LYP with a denser grid, three million calculations, was done.
With BO the total number of calculations is a few thousand, ${\cal O}(M \times N_{BO}) \approx 2500$, where $M$ is the number of DF calculations,  for a single evaluation of ${\cal L}$, and $N_{BO}$ is the number of iterations BO requires to find the minimum of ${\cal L}$; for the results presented here $M=39$, 7 atomic and 32 molecular calculations, and $N_{BO} \approx 60$.\\

We also studied the impact in the accuracy of the XC functionals for different basis sets; the values of $a_0$,  $a_X$, and  $a_C$ were optimized with BO.
We compared the results of PBE-PBE and B-LYP with the PBE0 and B3LYP functionals; Tables III-VIII SM.
We found that for all different basis set sizes the results predicted with the XC functionals, with  $a_0$,  $a_X$, and  $a_C$  optimized with BO, the accuracy of the predicted physical properties is higher.
Furthermore, we found that the accuracy of XC functionals with optimized parameters and smaller basis sets is still more accurate than standard XC functionals. 
For example, the predicted RMSE with PBE0/6-311++G($df, pd$) is 16\% larger than the one predicted with PBE-PBE/6-311G(d, p), Tables V and VI SM.
For the B3LYP/6-311++G($d, p$) functional, the predicted RMSE is 3.9\% larger than B-LYP/6-311++G($d, p$), Tables V and VI  SM.

Using larger basis sets, we found that PBE-PBE/6-311++G($df, pd$) is 15\% more accurate that PBE0/6-311++G($3df, 3pd$) for the atomization energies of the G1 data set, Table \ref{tab:ae_basisset}. 
However, the B-LYP functional with 6-311++G($df, pd$) basis set is only 1.3\% more accurate than the B3LYP functional with 6-311++G($3df, 3pd$), Table \ref{tab:ae_basisset}. 
For the results reported in Table \ref{tab:ae_basisset}, we optimized $a_0$,  $a_X$, and  $a_C$ by minimizing the RMSE of the atomization energies of the G1 molecules using BO with the UCB acquisition function with $\kappa = 0.1$. The molecular geometries used during the calculations were optimized in each step of the BO algorithm using each set of values of $a_0$, $a_X$, and $a_C$ proposed by the acquisition function at each iteration.

From Table \ref{tab:a0_ax_ac_coeff}, we can observe that the optimized values of $a_0$ and $a_X$ remained similar for basis sets with different sizes, but for both XC functionals the value of $a_C$ increased with the size of the basis set.  
For example, for the PBE-PBE functional $a_C$ changed from 0.0292 to 0.3537.
In the case of the B-LYP functional, the value of $a_C$ found by BO switched from 0.0201 to 0.7080. 
Additionally, the values of $a_0$,  $a_X$, and  $a_C$ of the B-LYP functional, optimized with 6-311++G($df,pd$), are similar to the values of $a_0$, $a_X$, and  $a_C$ of the B3LYP functional \cite{B_acm3_0}.\\

\begin{table}
\caption{\label{tab:a0_ax_ac_coeff}
Optimized values of $a_{0}$, $a_{X}$, and $a_{C}$ for the PBE-PBE and B-LYP functionals with BO for   different  basis sets.
We used BO with the UCB acquisition function to minimize the RMSE of the atomization energies of the G1 molecules \cite{G1,G1_1,G1_2}. 
We also report the values of $a_{0}$, $a_{X}$, and $a_{C}$ of the PBE0 and B3LYP functionals. }
\begin{ruledtabular}
\begin{tabular}{c | c | c c c  } 
$E_{XC}$ & basis set &  $a_{0}$ & $a_{X}$ & $a_{C}$\\ \hline \hline
PBE0\footnotemark[1] & -- & 0.25 & 0.75 & 1.0 \\
B3LYP\footnotemark[2]  & -- & 0.20 & 0.72 & 0.81 \\	\hline
\multirow{3}{*}{PBE-PBE\footnotemark[3]} & 6-31G($d$) & 0.1028 &  0.8708 &  0.0292  \\
					 & 6-311G($d,p$) & 0.1173 & 0.8742 & 0.1967 \\
					 & 6-311++G($df,pd$) & 0.1293 &  0.8855 & 0.3537  \\  \hline
\multirow{3}{*}{B-LYP\footnotemark[3]} & 6-31G($d$) & 0.1100 &  0.8293 &  0.0201 \\
					 & 6-311G($d,p$) &  0.1500 &  0.7658 &  0.3748  \\
					 & 6-311++G($df,pd$) & 0.1924 & 0.6962 & 0.7080 		 
\end{tabular}
\end{ruledtabular}
\footnotetext[1]{Refs. \cite{acm_theory,PBE0_1}}
\footnotetext[2]{Refs. \cite{B_acm3_0,B3LYP, B3LYP_2} }
\footnotetext[3]{Tables III SM and VIII SM }
\end{table}

\begin{table}
\tiny
\caption{\label{tab:ae_basisset} Theoretical and experimental atomization energies of the G1 molecules \cite{G1,G1_1,G1_2} predicted with different $E_{XC}$ functionals and two different basis sets.
We optimized the free parameters of PBE-PBE and B-LYP with BO using the UCB acquisition function with $\kappa=0.1$, Tables V-VIII SM. }
\begin{ruledtabular}
\begin{tabular}{c | c c | c c c c | c  } 
Molecule & PBE0\footnotemark[1] & B3LYP\footnotemark[1] & PBE-PBE\footnotemark[1]$^{,}$\footnotemark[3] & B-LYP\footnotemark[1]$^{,}$\footnotemark[4]  & PBE-PBE\footnotemark[2]$^{,}$\footnotemark[5] & B-LYP\footnotemark[2]$^{,}$\footnotemark[6] & Exp.\footnotemark[7]  \\ \hline \hline
H$_2$ &  98.00 &  103.84 &  107.51 &  107.06 &  105.59 &  104.26 &  103.5 \\ 
LiH &  50.72 &  56.39 &  59.52 &  59.06 &  57.84 &  56.76 &  56.0 \\ 
BeH &  52.90 &  55.05 &  56.26 &  56.56 &  55.63 &  55.25 &  46.9 \\ 
CH &  79.00 &  81.53 &  82.38 &  82.08 &  82.18 &  81.31 &  79.9 \\ 
CH$_2$(trip.) &  179.30 &  177.95 &  179.90 &  178.92 &  180.12 &  178.51 &  179.6 \\ 
CH$_2$(sing.) &  133.40 &  138.15 &  137.74 &  137.74 &  139.02 &  138.62 &  170.6 \\ 
CH$_3$ &  289.75 &  291.42 &  294.26 &  293.56 &  293.43 &  292.28 &  289.2 \\ 
CH$_4$ &  389.68 &  393.03 &  397.46 &  396.90 &  395.13 &  394.39 &  392.5 \\ 
NH &  80.63 &  83.53 &  83.27 &  83.08 &  83.81 &  82.96 &  79.0 \\ 
NH$_2$ &  171.25 &  176.27 &  175.35 &  175.31 &  176.16 &  175.33 &  170.0 \\ 
NH$_3$ &  273.60 &  279.80 &  278.64 &  278.99 &  279.91 &  279.76 &  276.7 \\ 
OH &  100.69 &  103.30 &  101.35 &  101.63 &  103.01 &  102.84 &  101.3 \\ 
OH$_2$ &  214.16 &  218.23 &  214.05 &  214.78 &  217.38 &  217.74 &  219.3 \\ 
FH &  131.65 &  134.11 &  129.68 &  130.31 &  133.46 &  133.88 &  135.2 \\ 
Li$_2$ &  19.06 &  20.48 &  22.44 &  20.78 &  21.84 &  20.55 &  24.0 \\ 
LiF &  130.91 &  135.84 &  134.65 &  135.00 &  134.99 &  135.04 &  137.6 \\ 
HCCH &  387.55 &  386.51 &  387.38 &  386.98 &  387.73 &  387.59 &  388.9 \\ 
H$_2$CCH$_2$ &  532.05 &  531.64 &  534.17 &  533.58 &  533.41 &  533.25 &  531.9 \\ 
H$_3$CCH$_3$ &  665.42 &  665.12 &  670.39 &  669.57 &  667.58 &  667.39 &  666.3 \\ 
CN &  176.13 &  176.68 &  178.19 &  177.13 &  178.30 &  176.48 &  176.6 \\ 
HCN &  301.45 &  303.66 &  304.24 &  304.17 &  304.09 &  303.73 &  301.8 \\ 
CO &  253.36 &  253.11 &  252.53 &  252.87 &  252.09 &  252.59 &  256.2 \\ 
HCO &  273.49 &  273.32 &  272.73 &  272.56 &  273.02 &  273.07 &  270.3 \\ 
H$_2$CO &  356.92 &  358.00 &  358.08 &  358.09 &  358.23 &  358.32 &  357.2 \\ 
H$_3$COH &  478.70 &  480.54 &  479.37 &  479.72 &  480.11 &  481.04 &  480.8 \\ 
N$_2$ &  221.97 &  226.12 &  225.23 &  225.41 &  225.03 &  224.45 &  225.1 \\ 
H$_2$NNH$_2$ &  402.04 &  408.60 &  407.09 &  407.67 &  408.01 &  408.49 &  405.4 \\ 
NO &  151.24 &  152.96 &  152.45 &  152.32 &  151.97 &  151.46 &  150.1 \\ 
O$_2$ &  122.36 &  121.68 &  121.73 &  121.19 &  120.03 &  119.74 &  118.0 \\ 
HOOH &  247.09 &  251.18 &  248.88 &  249.44 &  249.40 &  250.06 &  252.3 \\ 
F$_2$ &  32.63 &  34.76 &  35.99 &  35.20 &  33.32 &  32.87 &  36.9 \\ 
CO$_2$ &  386.40 &  382.39 &  380.60 &  381.04 &  380.72 &  381.61 &  381.9 \\ \hline
RMSE &  7.378 & 6.350 & 6.763 & 6.670 & 6.326 & 6.263 & \\
MAE &  3.845 & 3.067 & 3.853 &3.711 &  3.405 &  3.094 &
\end{tabular}
\end{ruledtabular}
\footnotetext[1]{6-311++G($df,pd$)}
\footnotetext[2]{6-311++G($3df,3pd$)}
\footnotetext[3]{$a_0$ = 0.1173,  $a_X$ = 0.8742, $a_C$ = 0.1967.}
\footnotetext[4]{$a_0$ = 0.1293,  $a_X$ = 0.8855, $a_C$ = 0.3537.} 
\footnotetext[5]{$a_0$ = 0.1500,  $a_X$ = 0.7658, $a_C$ = 0.3748.}
\footnotetext[6]{$a_0$ = 0.1924,  $a_X$ = 0.6962, $a_C$ = 0.7080.}
\footnotetext[7]{From Refs. \cite{G1,G1_1,G1_2}}
\end{table}

In the previous results, we used BO to minimize the RMSE of atomization energies to optimize the $a_0$, $a_X$, and $a_C$ of different XC functionals.
Here, we illustrate that BO can also optimize the free parameters of XC functionals for different physical properties, such as geometrical parameters.
We optimized $a_0$, $a_X$, and $a_C$ of 30 different XC functionals by minimizing the RMSE of the predicted geometrical parameters of the 15 diatomic molecules of the G1 data set \cite{G1,G1_1,G1_2}, Table \ref{tab:bl_g2}.\\

\begin{figure}
  \centering
	\includegraphics[width=0.45\textwidth]{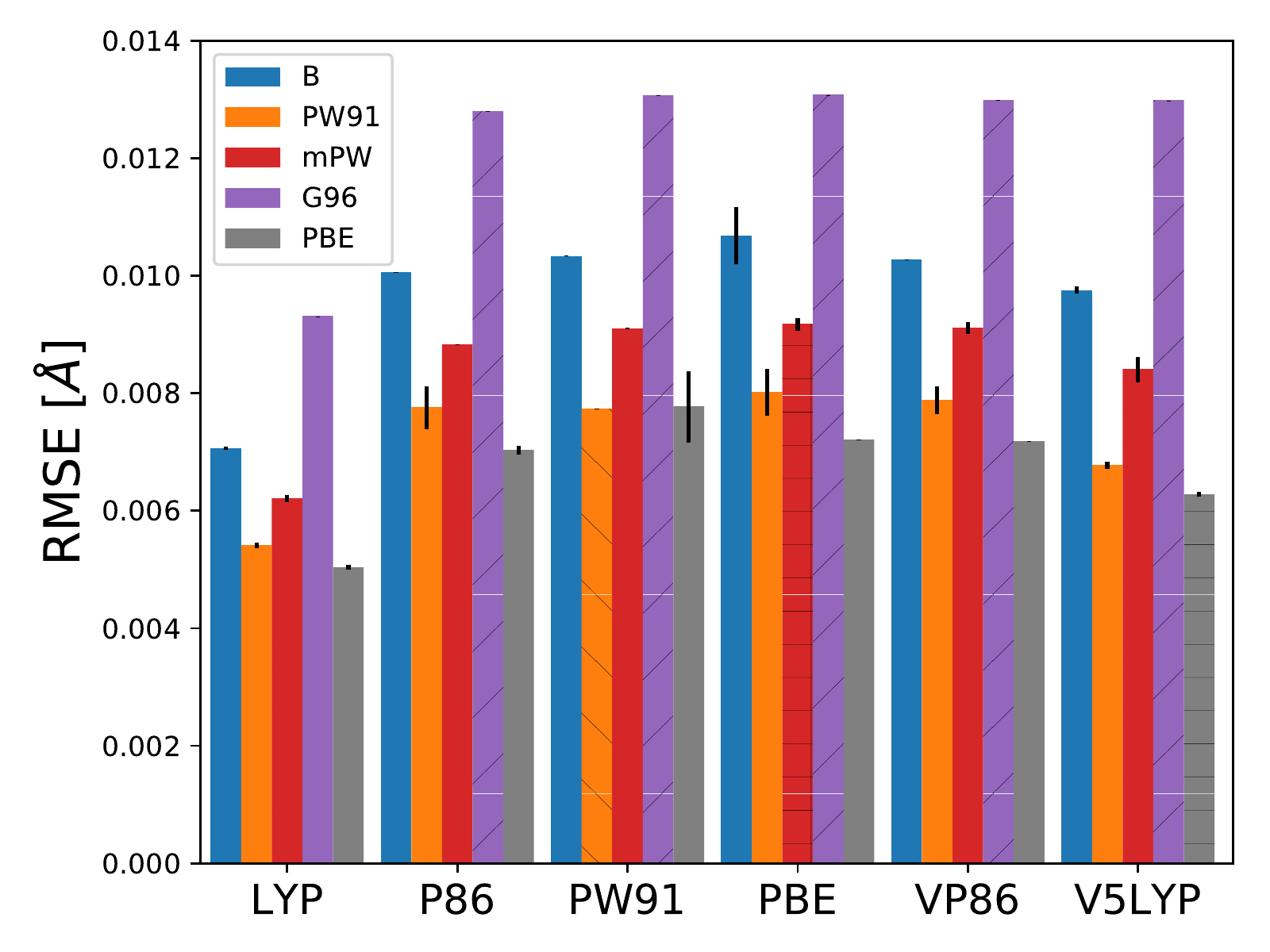}
	\includegraphics[width=0.45\textwidth]{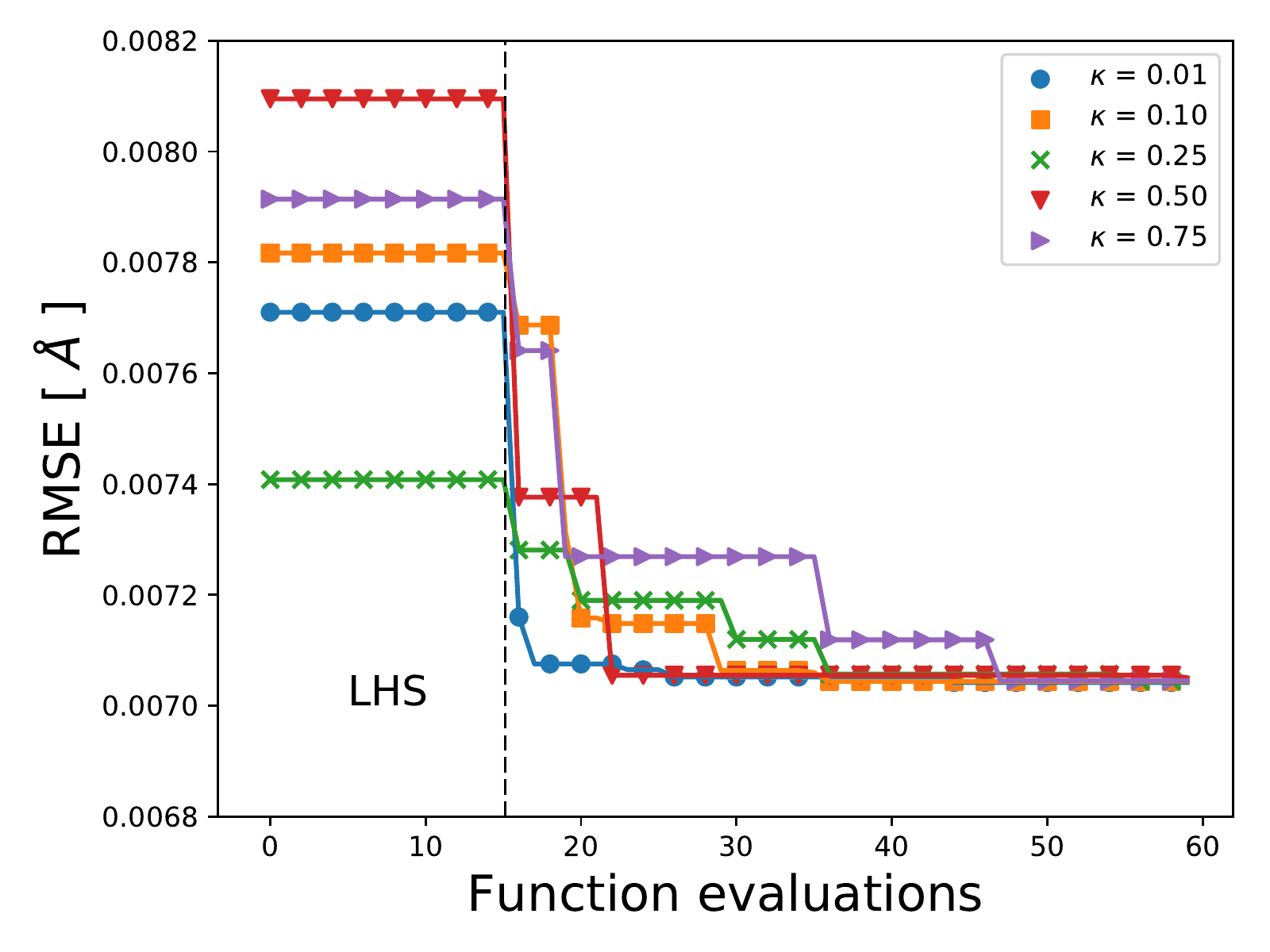}
	  \caption{(upper panel) Lowest RMSE, Eq. \ref{eqn:loss}, found by BO averaged over 3 different optimizations for 30 different XC functionals where $R_{m_i}$ is the bond length of the 15 diatomic molecules of the G1 molecules \cite{G1,G1_1,G1_2}.
	We considered 5 different $X$ functionals, colour and hatch coded in the inset of the graph, and the horizontal axis labels represent the 6 different $C$ functionals. 
    	For each XC functional, Eq. (\ref{eq:acm3}), we optimized the free parameters by initializing BO with 3 different set of initial points, $N_0 = 15$, sampled with LHS. For all optimizations, we use the UCB acquisition function with $\kappa = 0.1$. The maximum number of points allowed in the BO algorithm was 60, including the initial points.  
(lower panel) Lowest RMSE at each iteration of the BO algorithm for the optimization of the free parameters of Eq. (\ref{eq:acm3}) for the B-LYP functional. Each curve represents a different value of $\kappa$ used in the UCB acquisition function.}
	  \label{fig:bo_ucb_bl}
\end{figure}

For each XC functional we carried out 3 different optimizations, each with different 15 initial points sampled with the LHS algorithm \cite{LHS}. 
The maximum number of iterations allowed, for each optimization, was 60 total points including the LHS points. 
The averaged lowest RMSE found by BO for each XC functional is displayed in Fig. \ref{fig:bo_ucb_bl}. 
We use the UCB acquisition with $\kappa$ fixed to 0.1 and for all calculations we used the 6-311G($d,p$) basis set.
From Fig. \ref{fig:bo_ucb_bl} we can observe that any-value of $\kappa$ bellow 0.75 can find the minimum of the RMSE before 60 total evaluations.
The optimized coefficients for all 30 XC functionals are reported in Table IV SM.

To validate the accuracy of the optimized XC functionals we compare the results with two of the standards XC functionals, PBE0 and B3LYP. 
In Table  \ref{tab:bl_g2}, we reported the values of the bond length values of the 15 diatomic molecules of the G1 data set. 
We can observe that for both XC functionals the optimization of  $a_0$, $a_X$, and $a_C$ with BO yields to more accurate results.\\

\begin{table}
\caption{\label{tab:bl_g2}  Theoretical and experimental bond lengths [$\AA$] of G1 molecules \cite{G1,G1_1,G1_2} diatomic molecules. All values reported are obtained with 6-311G($d,p$).
The RMSE and MAE were computed with respect to the experimental bond lengths \cite{G1,G1_1,G1_2}.}
\begin{ruledtabular}
\begin{tabular}{c | c c c c | c  } 
Molecule & PBE0\footnotemark[1]  & B3LYP & B-LYP\footnotemark[2]$^{,}$\footnotemark[4] & PBE-PBE\footnotemark[3]$^{,}$\footnotemark[4] & Exp.\footnotemark[5]  \\ \hline \hline
H$_{2}$ &   0.745 &   0.744 &   0.741 &   0.738 &   0.742 \\ 
LiH &   1.597 &   1.592 &   1.588 &   1.587 &   1.595 \\ 
BeH &   1.347 &   1.344 &   1.339 &   1.338 &   1.343 \\ 
CH &   1.126 &   1.128 &   1.125 &   1.123 &   1.120 \\ 
NH &   1.040 &   1.044 &   1.042 &   1.040 &   1.045 \\ 
OH &   0.971 &   0.975 &   0.974 &   0.973 &   0.971 \\ 
FH &   0.916 &   0.920 &   0.919 &   0.918 &   0.917 \\ 
Li$_{2}$ &   2.705 &   2.705 &   2.694 &   2.694 &   2.670 \\ 
LiF &   1.559 &   1.560 &   1.565 &   1.566 &   1.564 \\ 
CN &   1.163 &   1.166 &   1.166 &   1.166 &   1.172 \\ 
CO &   1.125 &   1.127 &   1.127 &   1.127 &   1.128 \\ 
N$_{2}$ &   1.093 &   1.095 &   1.096 &   1.096 &   1.098 \\ 
NO &   1.143 &   1.148 &   1.149 &   1.149 &   1.151 \\ 
O$_{2}$ &   1.194 &   1.206 &   1.207 &   1.208 &   1.207 \\ 
F$_{2}$ &   1.401 &   1.408 &   1.412 &   1.416 &   1.412 \\ \hline
RMSE &  1.1E-2  &  9.68E-3  &  7.04E-3  &  7.21E-3  \\ 
MAE &  7.33E-3  &  5.30E-3  &  4.19E-3  &  4.67E-3  \\
\end{tabular}
\end{ruledtabular}
\footnotetext[1]{From Ref. \cite{acm_theory}}
\footnotetext[2]{$a_0$ = 0.1467,  $a_X$ =  0.6323 , $a_C$ = 0.001.}
\footnotetext[3]{$a_0$ =  0.1979,  $a_X$ = 0.7957 , $a_C$ = 0.001.}
\footnotetext[4]{Optimized with BO and $\alpha_{UCB}$ with $\kappa$ = 0.1.}
\footnotetext[5] {From Refs. \cite{G1,G1_1,G1_2}}
\end{table}

BO can also be applied to optimize range-separated density functionals \cite{LRDFT_0,LRDFT_1,LRDFT_2,LRDFT_3} using the \emph{mean absolute error} (MAE) function \cite{Reiher_BODFT}.
In Ref. \cite{ravh_thesis} we demonstrated that BO can optimize the parameter in the Yukawa potential between electrons,  commonly denoted as $\gamma$ \cite{LRDFT_0,LRDFT_1,LRDFT_2,LRDFT_3}.
We used the absolute difference between the highest occupied molecular orbital (HOMO) predicted with LCY-PBE and the ionization potential (IP) for the hydrogen molecule; 
\begin{eqnarray}
{\cal L}_{MAE} &=&  \left | \hat{E}^{HOMO}_{H_2}(\gamma) - IP \right |.\label{eqn:loss_mae} 
\end{eqnarray}
We demonstrated that, with a total of 6 points, the value of $\gamma$ for the LCY-PBE functional differed by 0.04 with respect to the reference one, $\hat{\gamma} = 1.2$, obtained with a brute-force search algorithm.
We used the UCB acquisition function with $\kappa=1$ for these calculations \cite{ravh_thesis}. \\

We also optimized the values of $a_0$, $a_X$, and $a_C$, for the PBE-PBE and B-LYP functionals by minimizing the MAE for the atomization energies of the G1 data set with BO. 
We used the molecular geometries optimized with MP2/6-31G($d$) and the UCB acquisition function with $\kappa=0.1$, \ref{eqn:a_ucb}. All energy calculations were done using 6-311G($d,p$).
The optimized free parameters of PBE-PBE and B-LYP functionals are reported in Table IX SM.
For the PBE-PBE functional,  $a_0 = 0.1231$, $a_X = 0.8977$, and $a_C = $ 0.4960 with MAE $= 3.398$ kcal mol$^{-1}$; and for B-LYP,  $a_0 = 0.1663 $, $a_X = 0.7233$, and $a_C = 0.7827$ with MAE $= 3.134$ kcal mol$^{-1}$. \\

\begin{figure}
  \centering
	\includegraphics[width=0.45\textwidth]{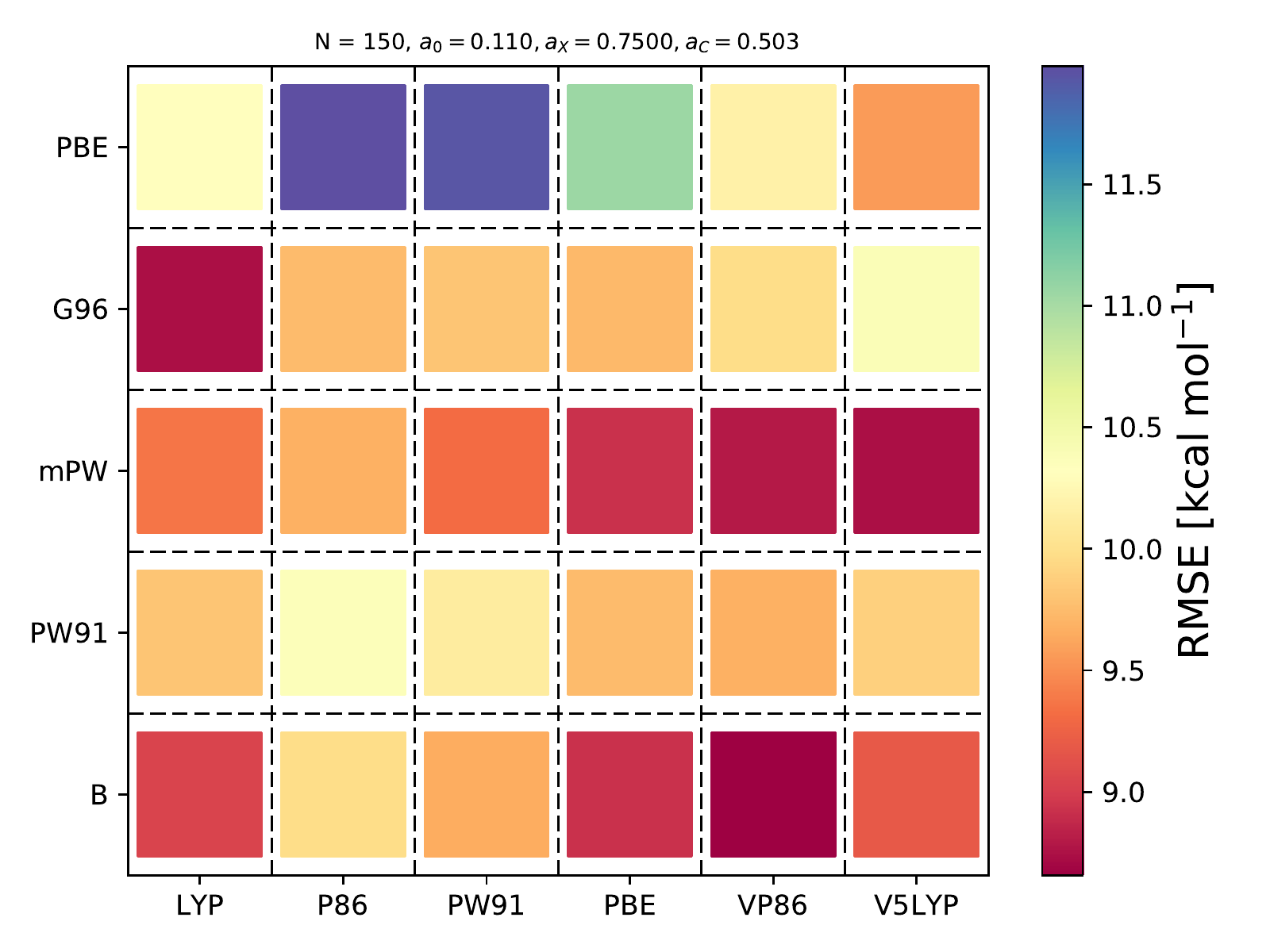}		
	\includegraphics[width=0.45\textwidth]{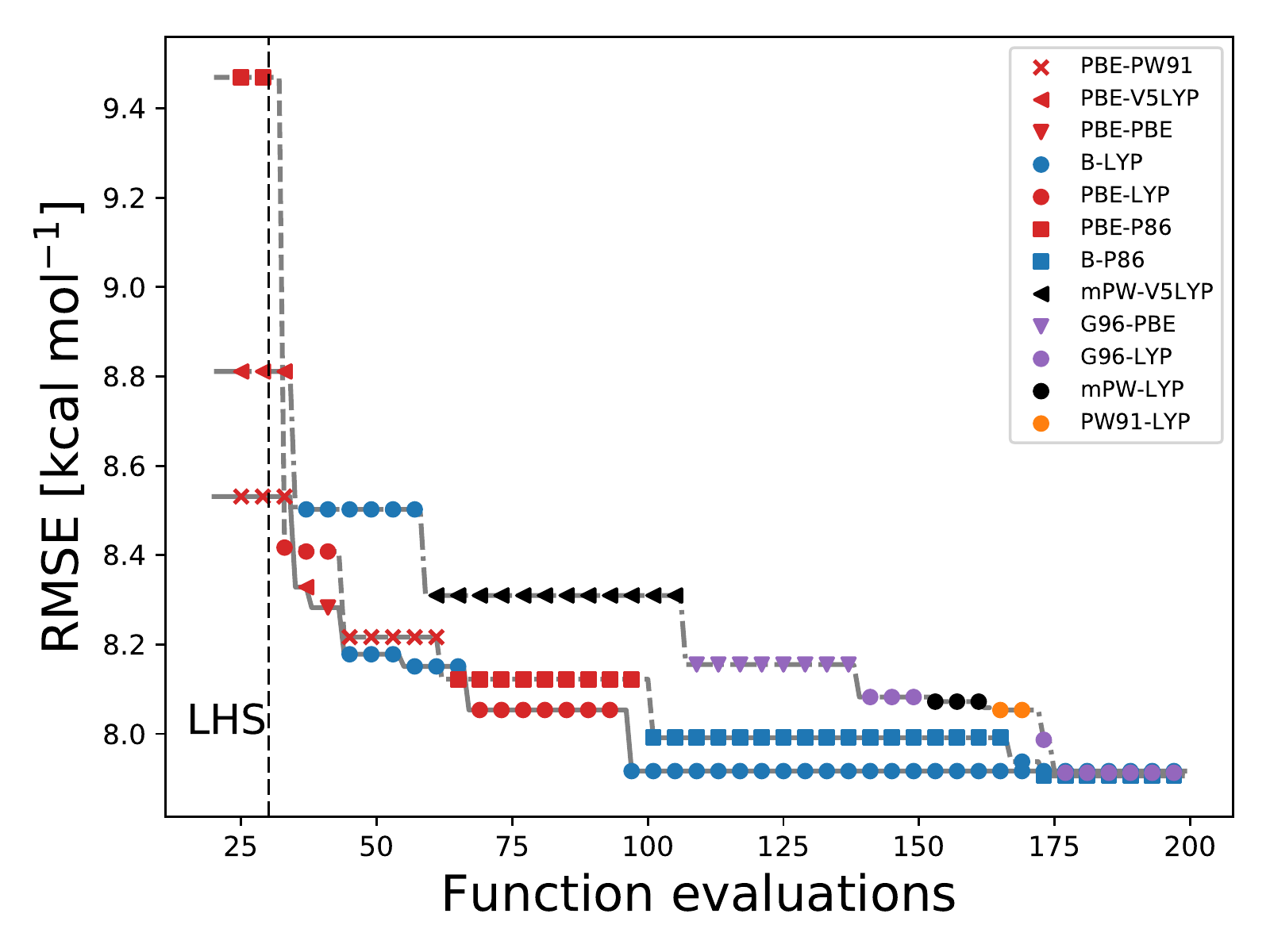}
	  \caption{(upper panel) GP prediction, trained with 150 points, of the RMSE for the 30 different DF models for $a_0, a_X$, and $a_C$ fixed to $0.11, 0.75, 0.503$ respectively.
	  (lower panel) The lowest RMSE found by BO at each iteration for 3 independent optimizations using the $\alpha_{UCB}$ with $\kappa=0.05$. 
	  The symbols represent the XC functional with the lowest RMSE selected by BO at each iteration. For each optimization we used 30 different initial points sampled with LHS.
	  For both panels the RMSE is computed using Eq. \ref{eqn:loss}, where $R_{m_i}$ are atomization energies of the G1 molecules \cite{G1,G1_1,G1_2}.  We used the floor function to integer round the variables that represent $E_X$ and $E_C$.}
	  \label{fig:bo_ucb_xcacm3}
\end{figure}

As it is known, some XC functionals tend to better describe some molecular systems than others. 
This has inspired multiple works where various DF models are compared  to each other using different benchmarks \cite{dft_HGordon_benchmark,DFT_benchmark0,DFT_benchmark1,DFT_benchmark2,select_dft}.
This can also be observed in Fig. \ref{fig:min_rmse_DFTall} where the RMSE is lower when the correlation functional is LYP or P86. 
Taking this into account, we wondered if BO could also help us identify the most appropriate XC functional.
From Eqs. (\ref{eqn:loss}--\ref{eqn:loss_mae}) it can be observed that ${\cal L}$ is a function of the DF model too, ${\cal L} = f({\cal M}_i,a_0,a_X,a_C)$.

Identifying the most appropriate DF model is also an optimization problem, 
\begin{eqnarray}
\mathbf{z^*},\mathbf{x^*} = \underset{\mathbf{z},\mathbf{x}}{\mathrm{arg\;min}} \; {\cal L}(\mathbf{z},\mathbf{x}), 
\end{eqnarray}
where $\mathbf{z}$ is an integer-valued vector, $z_i \in \mathbb{Z}^m$, used for labeling the different models. 
We define, $\mathbf{z} = [E_X,E_C]$ and $\mathbf{x} = [a_0,a_X,a_C]$.
We assigned an integer value to each $E_X$ \cite{exchange} and $E_C$ \cite{correlation}  functional, e.g., for PBE-PBE  $\mathbf{z} = [5,4]$ and for mPW-V5LYP $\mathbf{z} = [3,6]$.
Using BO we can efficiently navigate the DF space to select the optimal DF model for ${\cal L}(\mathbf{z},\mathbf{x})$ and by-pass the use of grid search methods \cite{sm_acq_floor}.
\\ 

In BO, the GP model is the surrogate model used to approximate ${\cal L}$. 
By including $E_X$ and $E_C$ in the feature space, the GP model learns the correlation between different XC models and $a_0$, $a_X$, and $a_C$, Fig. \ref{fig:bo_ucb_xcacm3}.
We also used the RBF kernel function, Eq. \ref{eqn:rbf_kernel}, in the GP model and the UCB acquisition function with $\kappa = 0.05$ \cite{sm_acq_dif_k}.  
The dimensionality of ${\cal L}$ changed to $d=5$; therefore increased the initial set of points, also sampled with LHS, to $N_0 = 30$, including random XC functionals. 
During the numerical optimization of the acquisition function we replaced the continuous values of $\mathbf{z}$ to the closest integer using the floor function \cite{sm_acq_floor}, e.g., $\lfloor{\mathbf{z}_i \rfloor} = \lfloor{[1.5,5.8]\rfloor}  = [1,5]$, which is B-VP86 \cite{exchange,correlation}. 
From Fig. \ref{fig:min_rmse_DFTall} we can observe that only 7 out of the 30 DF models considered here have an RMSE bellow 8.0 kcal mol$^{-1}$.
The goal, use BO to select the DF models with the lowest RMSE without any prior knowledge. 
Fig. \ref{fig:bo_ucb_xcacm3} illustrates that as the iterations of the algorithm progresses, BO samples different XC functionals to learn the most accurate combination of X and C functionals, including the optimal values of $a_0, a_C$ and $a_C$. 
Furthermore, BO is capable of selecting the DF model which RMSE is below 8.0 kcal mol$^{-1}$, Fig. \ref{fig:bo_ucb_xcacm3}.
We stressed that the BO algorithm learns the correlation between $a_0$, $a_X$, and $a_C$ with $E_X$ and $E_C$ to select the DF model with the lowest RMSE, Table XI SM.\\

\emph{Summary.} 
We have presented a powerful optimization method to calibrate DF models concerning a benchmark set of physical properties. 
BO algorithm relies on GP models to approximate ${\cal L}$ and an acquisition function to guide the sampling scheme towards the minimum of ${\cal L}$ without computing the gradient of ${\cal L}$. 
In this work, we illustrated that BO can optimize the free parameters of various DF models, e.g., hybrid-XC and range-separated functionals \cite{Reiher_BODFT,ravh_thesis}, for different type of loss functions, e.g., RMSE and MAE. 
This makes BO suitable also for optimizing many other computational physics and chemistry models \cite{BO_ff}, or DF models with a larger number of free parameters \cite{Roch_DFT}. 
Our results demonstrated that the optimization of DF models with BO is more efficiently than with grid search methods.
We also showed that the values of the free parameters of XC functionals could be optimized using low-computationally demanding calculations, e.g., small basis sets, and study systems with larger basis sets to increase the accuracy.

Due to the number of DF models currently available, the selection of DF models to accurately simulate a molecular system is also a computationally demanding task.
However, using BO algorithm one can efficiently navigate through the space of DFs. 
We demonstrated that BO can select the XC functional that better describes the system of interest while optimizing the free parameters of the DF model.
Our work illustrates the possibility to automate the selection and optimization of DF models to simulate physical properties more accurately.\\

We acknowledge useful discussions with R. V. Krems and R. Ghassemizadeh, and M. Rueda-Becerril for constructive criticism of the manuscript.
This work was supported by the Natural Sciences and Engineering Research Council of Canada (NSERC).

\bibliography{reference}

\begin{thebibliography}{99}

\bibitem{dftbook}
Wolfram Koch, and Max C. Holthausen,
{\it A chemist's guide to density functional theory} (Wiley-VCH, 2000).

\bibitem{dft_revmodphys}
R. O. Jones,
\emph{Rev. Mod. Phys.} {\bf 87}, 897 (2015).

\bibitem{dft_HGordon_benchmark}
N. Mardirossian, and M. Head-Gordon, 
\emph{Mol. Phys.} {\bf 115}, 2315 (2017). 

\bibitem{DFT_benchmark0}
S. Kurth, J.P. Perdew, and P. Blaha, 
\emph{Int. J. Quantum Chem.} {\bf 69}, 75, 889 (1999)

\bibitem{DFT_benchmark1}
V. N. Staroverov, G. E. Scuseria, J. Tao, and J. P. Perdew,
\emph{J. Chem. Phys.} {\bf 119}, 12 129 (2003).

\bibitem{DFT_benchmark2}
V. N. Staroverov, G. E. Scuseria, J. Tao, and J. P. Perdew,
\emph{Phys. Rev. B} {\bf 69}, 075102 (2004).

\bibitem{DFT_to_ML}
G. R. Schleder, A. C. M. Padilha,  C. M. Acosta, M.Costa, and A. Fazzio,
\emph{J. Phys. Materials} (2019).

\bibitem{ML_optbook}
S. Sra, S. Nowozin, and S. J. Wright,
{\it  Optimization for Machine Learning} (The MIT Press, Cambridge, 2012).
\bibitem{ML_opt}
F. E. Curtis, and K. Scheinberg,
 arXiv:1706.10207 

\bibitem{B_acm3_0}
A. D. Becke, \emph{J. Chem. Phys.} {\bf 98}, 5648 (1993).

\bibitem{B_acm3_1}
A. D. Becke, \emph{J. Chem. Phys.} {\bf 98}, 1372 (1993).

\bibitem{OptB3LYP}
L. Lu,
\emph{ Int. J. Quantum Chem.} {\bf 115}, 471 (2015).

\bibitem{vonLilienfeld}
L.-F. Arsenault, A. Lopez-Bezanilla, O. A. von Lilienfeld, and A. J. Millis,
{\it Phys. Rev. B} {\bf 90}, 155136 (2014).

\bibitem{vonLilienfeld_2}
L.-F.  Arsenault,  O.  A.  von  Lilienfeld,  and  A.  J.  Millis,
arXiv:1506.08858.

\bibitem{Aspuru_NNmolecules}
D. Duvenaud, D. Maclaurin, J. Aguilera-Iparraguirre, R. G\'omez-Bombarelli, T. Hirzel, A. Aspuru-Guzik, and R. P. Adams
{\it Adv. Neural Inf. Process. Syst.} {\bf 28}, 2224 (2015).

\bibitem{Aspuru_NNmolecules_2}
R. G\'omez-Bombarelli,  J. N. Wei,  D. Duvenaud, J. M. Hern\'andez-Lobato,  B. S\'anchez-Lengeling,  D. Sheberla, J. Aguilera-Iparraguirre,  T. D.  Hirzel, R. P. Adams,  and A. Aspuru-Guzik, 
{\it ACS Cent. Sci.} {\bf 4}, 268 (2018).

\bibitem{Wang_prb}
L. Wang,
{\it Phys. Rev. B} {\bf 94}, 195105 (2016).

\bibitem{Carleo_science}
G. Carleo, and M. Troyer,
{\it Science} {\bf 355}, 602 (2017).

\bibitem{Carrasquilla_natphys}
J. Carrasquilla, and R. G. Melko,
{\it Nat. Phys.} {\bf 13}, 431 (2017).

\bibitem{ravh_prl}
R. A. Vargas-Hern\'andez, J. Sous, M. Berciu, and R. V. Krems,
{\it Phys. Rev. Lett.} {\bf 121}, 255702 (2018).

\bibitem{KE_DFT}
J. C. Snyder, M. Rupp, K. Hansen, K.-R. M\"uller, and K. Burke
{\it Phys. Rev. Lett.} {\bf 108}, 253002  (2012).

\bibitem{ML_DFT}
F. Brockherde, L. Vogt, L. Li, M. E. Tuckerman, K. Burke, and K.-R. M\"uller, 
{\emph Nat. Comms.} {\bf 8}, 872 (2017).

\bibitem{ML_DFT_2}
M. Bogojeski, F. Brockherde, L. Vogt-Maranto, L. Li, M. .  Tuckerman, K. Burke, and K.-R. M\"uller,
arXiv:1811.06255

\bibitem{RK_bayesML}
R. V. Krems,
arXiv:1904.03730

\bibitem{GPBEC}
P. B. Wigley, P. J. Everitt, A. van den Hengel, J. W. Bastian, M. A. Sooriyabandara, G. D. McDonald, K. S. Hardman, C. D. Quinlivan, P. Manju, C. C. N. Kuhn, I. R. Petersen, A. N. Luiten, J. J. Hope, N. P. Robins, and M. R. Hush,
{\it Scientific Reports}  {\bf 6}, 25890 (2016).

\bibitem{Reiher_chemrxn_error}
G. N. Simm, and M. Reiher,
\emph{J. Chem. Theory Comput.} {\bf 12},  2762 (2016).

\bibitem{Reiher_chemrxn_error2}
G. N. Simm, and M. Reiher,
\emph{J. Chem. Theory Comput.} {\bf 14}, 5238 (2018).

\bibitem{phoenics}
F. H\"ase, L. M. Roch, C. Kreisbeck, and A. Aspuru-Guzik,
{\it ACS Cent. Sci.}  {\bf 4}, 1134 (2018).

\bibitem{chemos}
L. M. Roch, F. H\"ase, C. Kreisbeck, T. Tamayo-Mendoza, L. P.-E. Yunker, J. E Hein, and A. Aspuru-Guzik,
{\it Science Robotics} {\bf 3}, 19 (2018).

\bibitem{BEE_PRL}
J. J. Mortensen, K. Kaasbjerg, S. L. Frederiksen, J. K. Nørskov, J. P. Sethna, and K. W. Jacobsen
{\it Phys. Rev. Lett.} {\bf 95}, 216401 (2005).

\bibitem{BEE_PRB}
J. Wellendorff, K. T. Lundgaard, A. M\o{}gelh\o{}j, V. Petzold, D. D. Landis, J. K. N\o{}rskov, T. Bligaard, and K. W. Jacobsen, 
{\it Phys. Rev. B} {\bf 85}, 235149 (2012)

\bibitem{Bayes_DFT_0}
A. J. Medford, J. Wellendorff, A. Vojvodic, F. Studt, F. Abild-Pedersen, K. W. Jacobsen, T. Bligaard, and J. K. Norskov, 
\emph{Science} {\bf 345}, 6193 (2014). 

\bibitem{mBEEF}
J. Wellendorff, K. T. Lundgaard, K. W. Jacobsen, and T. Bligaard,
\emph{J. Chem.  Phys. } {\bf 140},  144107  (2014).

\bibitem{Walter_BEE}
R. W\"urdemann, H. H. Kristoffersen, M. Moseler, and M. Walter,
\emph{ J. Chem. Phys. } {\bf 142}, 124316 (2015).

\bibitem{Bayes_DFT_1}
M. Aldegunde, J. R. Kermode, and N. Zabaras, 
\emph{J. Comput. Phys.} {\bf 311}, 173 (2016).

\bibitem{Bayes_DFT_water}
M. Fritz, M. Fern\'andez-Serra, and J. M. Soler
\emph{ J. Chem. Phys. } {\bf 144}, 224101 (2016).

\bibitem{BO_Adams}
 J. Snoek, H. Larochelle, and R. P. Adams, 
 \emph{Adv. Neur. Inf. Process. Sys.} {\bf 25}, 2951(2012).

\bibitem{BO_Freitas}
B. Shahriari, K. Swersky, Z. Wang, R .P. Adams, and N. de Freitas, \emph{Proc. IEEE} {\bf 104}, 148 (2016).

\bibitem{BO_PES}
S. Carr, R. Garnett, and C. Lo,
``BASC: Applying Bayesian Optimization to the Search for Global Minima on Potential Energy Surfaces'',
in Proceedings of The 33rd International Conference on Machine Learning, (PMLR, 2016), pp. 898-907. 

\bibitem{BO_geo}
L. Chan, G. Hutchison, and G. Morris,
chemrxiv.7228940.v1

\bibitem{ravh_njp}
 R. A. Vargas-Herná\'andez, Y. Guan, D. H. Zhang,  and R. V. Krems,
{\it New J.  Phys.} {\bf 21}, 022001(2019).

\bibitem{BO_mat}
T. Ueno, T. D. Rhone, Z. Hou, T. Mizoguchi, and K. Tsuda,
{\it Mater. Discovery} {\bf 4}, 18 (2016).

\bibitem{BO_dft_calc}
R. Jalem, K. Kanamori, I. Takeuchi, M. Nakayama, H. Yamasaki, and T. Saito,
{\it Sci. Rep.} {\bf 8}, 5845 (2018).

\bibitem{BO_phonono_transport}
Shenghong Ju, Takuma Shiga, Lei Feng, Zhufeng Hou, Koji Tsuda, and Junichiro Shiomi
{\it Phys. Rev. X} {\bf 7}, 021024 (2017).

\bibitem{BO_ising}
R. Tamura, and  K. Hukushima,
{\it PLoS ONE} {\bf 13}, e0193785 (2018).

\bibitem{gpbook}
C. E. Rasmussen, and C. K. I. Williams, 
{\it Gaussian Processes for Machine Learning} (The MIT Press, Cambridge, 2006).

\bibitem{sm_gp}
 See Supplemental Material ``URL will be inserted by publisher'' section II.A.

\bibitem{G09}
M. J. Frisch, G. W. Trucks, H. B. Schlegel, G. E. Scuseria, M. A. Robb, J. R.
Cheeseman, G. Scalmani, V. Barone, B. Mennucci, G. A. Petersson, H. Nakatsuji, M. Caricato, X. Li, H. P. Hratchian, A. F. Izmaylov, J. Bloino, G. Zheng, J. L. Sonnenberg, M. Hada, M. Ehara, K. Toyota, R. Fukuda, J. Hasegawa, M. Ishida, T. Nakajima, Y. Honda, O. Kitao, H. Nakai, T. Vreven, J. A. Montgomery, Jr., J. E. Peralta, F. Ogliaro, M. Bearpark, J. J. Heyd, E. Brothers, K. N. Kudin, V. N. Staroverov, R. Kobayashi, J. Normand, K. Raghavachari, A. Rendell, J. C. Burant, S. S. Iyengar, J. Tomasi, M. Cossi, N. Rega, J. M. Millam, M. Klene, J. E. Knox, J. B. Cross, V. Bakken, C. Adamo, J. Jaramillo, R. Gomperts, R. E. Stratmann, O. Yazyev, A. J. Austin, R. Cammi, C. Pomelli, J. W. Ochterski, R. L. Martin, K. Morokuma, V. G. Zakrzewski, G. A. Voth, P. Salvador, J. J. Dannenberg, S. Dapprich, A. D. Daniels, O. Farkas, J. B. Foresman, J. V. Ortiz, J. Cioslowski, D. J. Fox, Gaussian 09, Revision A.02, Gaussian: Wallingford CT, 2009.

\bibitem{G1}
J.  A.  Pople,  M.  Head-Gordon,  D.  J.  Fox,  K.  Raghavachari,  and  L.  A.  Curtiss, 
\emph{J.  Chem.  Phys. } {\bf 90},  5622 (1989). 

\bibitem{G1_1}
L.  A.  Curtiss, C. Jones,  G.  w.  Trucks, K.  Raghavachari, and J.  A.  Pople,
\emph{J. Chem.  Phys. } {\bf 93},  2537 (1990).

\bibitem{G1_2}
D. Feller, and K. A. Peterson,
\emph{J. Chem.  Phys. } {\bf 110}, 8384 (1999).

\bibitem{acm_theory0}
J. P. Perdew, M. Ernzerhof, and K. Burke,
\emph{ J. Chem. Phys. } {\bf 105}, 9982 (1996).

\bibitem{acm_theory}
C. Adamo, and V. Barone,
\emph{ J. Chem. Phys.} {\bf 110}, 6158 (1999).

\bibitem{PBE0_1}
M. Ernzerhof, and G. E. Scuseria,
\emph{J. Chem. Phys.} {\bf 110}, 5029 (1999).

\bibitem{SM_PBE0}
 See Supplemental Material ``URL will be inserted by publisher'' section II.

\bibitem{exchange}
Exchange functionals, $E_X = [$ B, PW91, mPW, G96, PBE $]$

\bibitem{correlation}
Correlation functionals, $E_C = [$ LYP, P86, PW91, VP86, V5LYP $]$

\bibitem{LHS}
M. Stein,
\emph{Technometrics} {\bf 29}, 143 (1987).


\bibitem{SM_acm3_tab}
 See Supplemental Material ``URL will be inserted by publisher'' Table II.

\bibitem{SM_acm3_bl_tab}
 See Supplemental Material ``URL will be inserted by publisher'' Table III.


\bibitem{B3LYP}
K. Kim, and K. D. Jordan,
{\it J. Phys. Chem.} {\bf 98}, 10089 (1994).

\bibitem{B3LYP_2}
P.J. Stephens; F. J. Devlin; C. F. Chabalowski; M. J. Frisch,
{\it J. Phys. Chem.} {\bf 98}, 11623 (1994). 
 
\bibitem{LRDFT_0}
T. Leininger, H. Stoll, H.-J. Werner, and A. Savin,
\emph{Chem. Phys. Let.} {\bf 275}, 151  (1997).

\bibitem{LRDFT_1}
H. Iikura, T. Tsuneda, T. Yanai, and K. Hirao,
\emph{J. Chem.  Phys. } {\bf 115},  3540 (2001).

\bibitem{LRDFT_2}
Y. Akinaga, and S. Ten-no,
 Range-separation by the Yukawa potential in long-range corrected density functional theory with Gaussian-type basis functions. \emph{Chem. Phys. Let.} {\bf 462}, 348  (2008).

\bibitem{LRDFT_3}
M. Seth, and T. Ziegler,
\emph{J. Chem. Theory Comput.} {\bf 8}, 901 (2012).

\bibitem{Reiher_BODFT}
J. Proppe, S. Gugler, and M. Reiher, 
arXiv:1906.09342 

\bibitem{ravh_thesis}
Vargas-Herná\'andez, R. A. (2018). Applications of machine learning for solving complex quantum problems (T). PhD Thesis University of British Columbia.

\bibitem{select_dft}
D. Rappoport, N. R. Crawford, F. Furche and K. Burke,
{\it Approximate Density Functionals: Which Should I Choose?.}
In Encyclopedia of Inorganic Chemistry (2009). 


\bibitem{sm_acq_floor}
See Supplemental Material ``URL will be inserted by publisher''  Table III for comparing the BO results when the floor function was used  and not during the optimization of the acquisition function.  

 \bibitem{sm_acq_dif_k}
See Supplemental Material ``URL will be inserted by publisher''  Fig. SM 3 and Table IV  for the BO results when different values of $\kappa$ where used.

\bibitem{BO_ff}
R. B. Jadrich, B. A. Lindquist, and T. M. Truskett,
 arXiv:1706.05405

\bibitem{Roch_DFT}
L. M. Roch, and K. K. Baldridge, 
\emph{Phys. Chem. Chem. Phys.} {\bf 19}, 26191 (2017).




\end{thebibliography}

\end{document}



\title{Supplemental Material for\\
``Bayesian optimization for calibrating and selecting hybrid-density functional models"
}



\author{Rodrigo A. Vargas-Hern\'andez}
\address{Department of Chemistry, University of British Columbia, Vancouver, British Columbia, Canada, V6T 1Z1. \\
Current Address: Chemical Physics Theory Group, Department of Chemistry, and Center for Quantum Information and Quantum Control, University of Toronto,Toronto, Ontario, M5S 3H6, Canada.
}


\date{\today}

\begin{abstract}

\end{abstract}

\maketitle

The purpose of this supplemental material is to provide details of the numerical calculations we present in this work and explain the Bayesian optimization algorithm. 
In section I we present the details of the quantum chemistry calculations, and in section II we introduced introduce Gaussian Process (GP) models and the Bayesian optimization (BO) algorithm.
Section III contains all the results presented in this work.


\section{hybrid-Density functionals}
Density functionals (DF) methods are widely use in different fields such as biology, material science and theoretical chemistry. 
The most common DF methods are the hybrid-density functionals \cite{B_acm3_0,B_acm3_1}, 
\begin{eqnarray}
E^{acm3}_{XC} &=& E^{LSD}_{XC} + a_0(E^{exact}_{X} - E^{LSD}_{X}) + a_X(E^{GGA}_{X} - E^{LSD}_{X}) + a_C(E^{GGA}_{C} - E^{LSD}_{C}),
\label{eq:acm3}
\end{eqnarray}
where $a_0$, $a_X$ and $a_C$ are adjustable parameters,  $E^{GGA}_{X}$ and $E^{GGA}_{C}$ are the generalized gradient approximation (GGA) exchange and  correlation, and $E^{LSD}$ is the local spin density (LSD) part.  
$a_0$ and $a_X$ combine local and non-local treatments of exchange (X) and $a_C$ combines the correlation (C) between LSD and GGA functionals.

The optimization of DF is usually carried by tuning the parameters of DF models that best reproduce experimental data. 
In this work we considered the \emph{root mean square error} (RMSE) function with respect to the atomization energies of the Gaussian-1 (G1) database \cite{G1,G1_1,G1_2}, 
\begin{eqnarray}
f = \sqrt{\frac{1}{|M|}\sum^{|M|}_{m_i} \left(R^{DFT}_{m_i} - R^{exact}_{m_i} \right )^2},
\label{eqn:loss_dft}
\end{eqnarray}
where $|M|$ is 32 atomization energies considered and $R^{exact}_{m_i}$ is the experimental atomization energy for molecule $m_i$ reported in Table \ref{tab:ae_g1}. $R^{DFT}_{m_i}$ is the atomization energy predicted with a DF model; $R^{DFT}_{m_i}$ is a function of the DF parameters, $ R^{DFT}_{m_i} = f(a_0,a_X,a_C)$.\\
For all calculations the molecular geometries used in the DF calculations were optimized with MP2/6-31G(d), and for all DF calculations we used 6-31G(d).
All calculations were performed with the Gaussian 09 suite \cite{G09}.


\begin{table}[ht]
\caption{Atomization energies $D_0$ [kcal mol$^{-1}$] \cite{g1_ae}.}
\begin{minipage}[b]{0.3\linewidth}\centering
\begin{tabular}{l |c }
 & Exp.   \\ \hline \hline 
H$_2$ &  103.5     \\
LiH & 56.0     \\
BeH & 46.9     \\
CH & 79.9   \\
CH$_2$ (trip.) & 179.6    \\
CH$_2$ (sing.)  & 170.6    \\
CH$_3$ & 289.2    \\
CH$_4$ & 392.5   \\
NH & 79.0    \\
NH$_2$  & 170.0   \\ 
NH$_3$  & 276.7   \\ 
OH  & 101.3    \\     
H$_2$O  & 219.3   \\ 
HF  & 135.2    \\ 
Li$_2$  & 24.0   \\ 
LiF  & 137.6  
\end{tabular}
\end{minipage}
\hspace{0.5cm}
\begin{minipage}[b]{0.3\linewidth}
\begin{tabular}{l |c }
 & Exp.   \\ \hline \hline 
 C$_2$H$_2$  & 388.9    \\ 
C$_2$H$_4$  & 531.9   \\
C$_2$H$_6$  & 666.3    \\
  CN & 176.6    \\ 
  HCN & 301.8   \\
  CO & 256.2   \\  
  HCO & 270.3   \\  
  H$_2$CO & 357.2    \\  
  CH$_3$OH & 480.8   \\  
  N$_2$ & 255.1   \\  
  N$_2$H$_4$ & 405.4   \\  
  NO & 150.1    \\  
  O$_2$ & 118.0   \\  
  H$_2$O$_2$ & 252.3    \\  
  F$_2$& 36.9    \\  
  CO$_2$ & 381.9   
\end{tabular}
\end{minipage}
\label{tab:ae_g1}
\end{table}

\begin{table}[ht]
\caption{Experimental bond lengths for diatomic molecules in G1 data set [$\AA$] \cite{g1_ae}.}
\begin{tabular}{l |c }
 & Exp.   \\ \hline \hline 
H$_{2}$ &    0.742 \\ 
LiH &    1.595 \\ 
BeH &    1.343 \\ 
CH &   1.120 \\ 
NH &     1.045 \\ 
OH &     0.971 \\ 
FH &    0.917 \\ 
Li$_{2}$ &   2.670 \\ 
LiF &   1.564 \\ 
CN &  1.172 \\ 
CO &   1.128 \\ 
N$_{2}$ & 1.098 \\ 
NO &    1.151 \\ 
O$_{2}$ &     1.207 \\ 
F$_{2}$ &    1.412 
\end{tabular}
\label{tab:bl_g1}
\end{table}

\break
\section{Gaussian process and Bayesian optimization}

Bayesian optimization is a sequential search algorithm designed to find the minimum/maximum of a non-analytic function, denoted here as $f$, without requiring the gradient of the function \cite{BO_Adams, BO_Freitas},
\begin{eqnarray}
\mathbf{x^*} = \underset{\mathbf{x}}{\mathrm{arg\;min}}  f(\mathbf{x}).
\end{eqnarray}
BO is constructed using two ingredients, a probabilistic model that approximates $f$ and an acquisition function that quantifies the informational gain if $f$ where to evaluated in a new point. 
Here we denoted ${\cal F}$ as the probabilistic model.
\subsection{Gaussian process}
In this work we used Gaussian process (GP) models as the probabilistic model that approximates $f$, ${\cal F} \approx f$.
GP models are robust supervised learning models that assume the data is Gaussian distributed \cite{gpbook}. 
The prediction of a GP model is a normal distribution characterized by a mean $\mu(\cdot)$ and a standard deviation $\sigma(\cdot)$, given as
\begin{eqnarray}
\mu(\bm x_*) &=& K({\bm x_*},X)^\top \left [ K(X, X) + \sigma_n^2 I \right ] ^{-1}{\bm y} \label{eq:gp_mu}\\
\sigma(\bm x_*) &=& K({\bm x_*},{\bm x_*}) - K({\bm x_*},X)^\top\left [ K(X,X) + \sigma_n^2 I\right ]^{-1}K({\bm x_*},X) \label{eq:gp_std} ,
\end{eqnarray} 
where $K$ is the design or covariance matrix and its matrix elements are computed using a kernel function, $K_{i,j} = k(\mathbf{x}_i,\mathbf{x}_i)$. For this work we only considered the \emph{radial basis function} (RBF) kernel,
 \begin{eqnarray}
k_{RBF}(\mathbf{x}_i,\mathbf{x}_j) = \exp\left(-\frac{1}{2}(\mathbf{x}_i-\mathbf{x}_j)^\top M (\mathbf{x}_i-\mathbf{x}_j)\right),
\label{eqn:rbf_kernel}
\end{eqnarray}
where
\begin{eqnarray}
M = \begin{pmatrix}
\ell_1 &  & 0 \\ 
 &\ddots   & \\ 
0 &  & \ell_d
\end{pmatrix},
\end{eqnarray}
and $\ell_i$ is the length-scale parameter for each dimension of $\mathbf{x}$. We described all  $\ell_d$s jointly as denoted as $\mathbf{\theta}$, $\mathbf{\theta} = [\ell_1, \cdots, \ell_d]$. The parameters of the kernel function, $\mathbf{\theta}$, are optimized by maximizing the log marginal likelihood,
 \begin{eqnarray}
\log p(\mathbf{y}|X,\mathbf{\theta}) &=& -\frac{1}{2}\mathbf{y}^\top K(X,X)^{-1} \mathbf{y}  -\frac{1}{2}\log|K(X,X)| - \frac{N}{2}\log(2\pi),
\end{eqnarray}
where $N$ is the total number of points in ${\cal D}$ and $|K(X,X)|$ is the determinant of the design matrix. 

\subsection{Acquisition function}
The acquisition function, denoted as $\alpha$, is the function that guides the sampling scheme in BO towards the minimum/maximum of $f$. 
An acquisition function can only be constructed using the information obtained from ${\cal F}$.
There are several acquisition functions, in this work we only used two; the expected improvement (EI) and the upper confidence bound  (UCB),
\begin{eqnarray}
\alpha_{EI}(\mathbf{x}) &=&( \mu(\mathbf{x}) - y_{max})\Phi(z(\mathbf{x};y_{max})) + \sigma(\mathbf{x})\phi(z(\mathbf{x};y_{max}))\label{eqn:a_ei} \\
\alpha_{UCB}(\mathbf{x}) &=& \mu(\mathbf{x})+ \kappa \sigma(\mathbf{x}), \label{eqn:a_ucb}
\end{eqnarray}
where $\mu(\mathbf{x})$ and $\sigma(\mathbf{x})$ are the mean and the standard deviation from a GP model, Eqs. (\ref{eq:gp_mu}-\ref{eq:gp_std}). For EI, $\Phi(\cdot)$  is the normal cumulative distribution and $\phi(\cdot)$ is the normal probability distribution,$z(\mathbf{x};y_{min}) = (\mu(\mathbf{x})-y_{min})/ \sigma(\mathbf{x})$, and $y_{min}$ is the minimum value observed in the training data, $y_{min} = \text{arg\;min }\mathbf{y}$. The UCB function has a hyper-parameter $\kappa$ known as the exploration-exploitation constant. $\kappa$ controls the sampling scheme of $\alpha_{UCB}$ towards exploratory moves or exploitation moves. 
 
\subsection{Bayesian Optimization}
\begin{figure}[h]
    \centering
    \begin{subfigure}[b]{0.42\textwidth}
        \includegraphics[width=\textwidth]{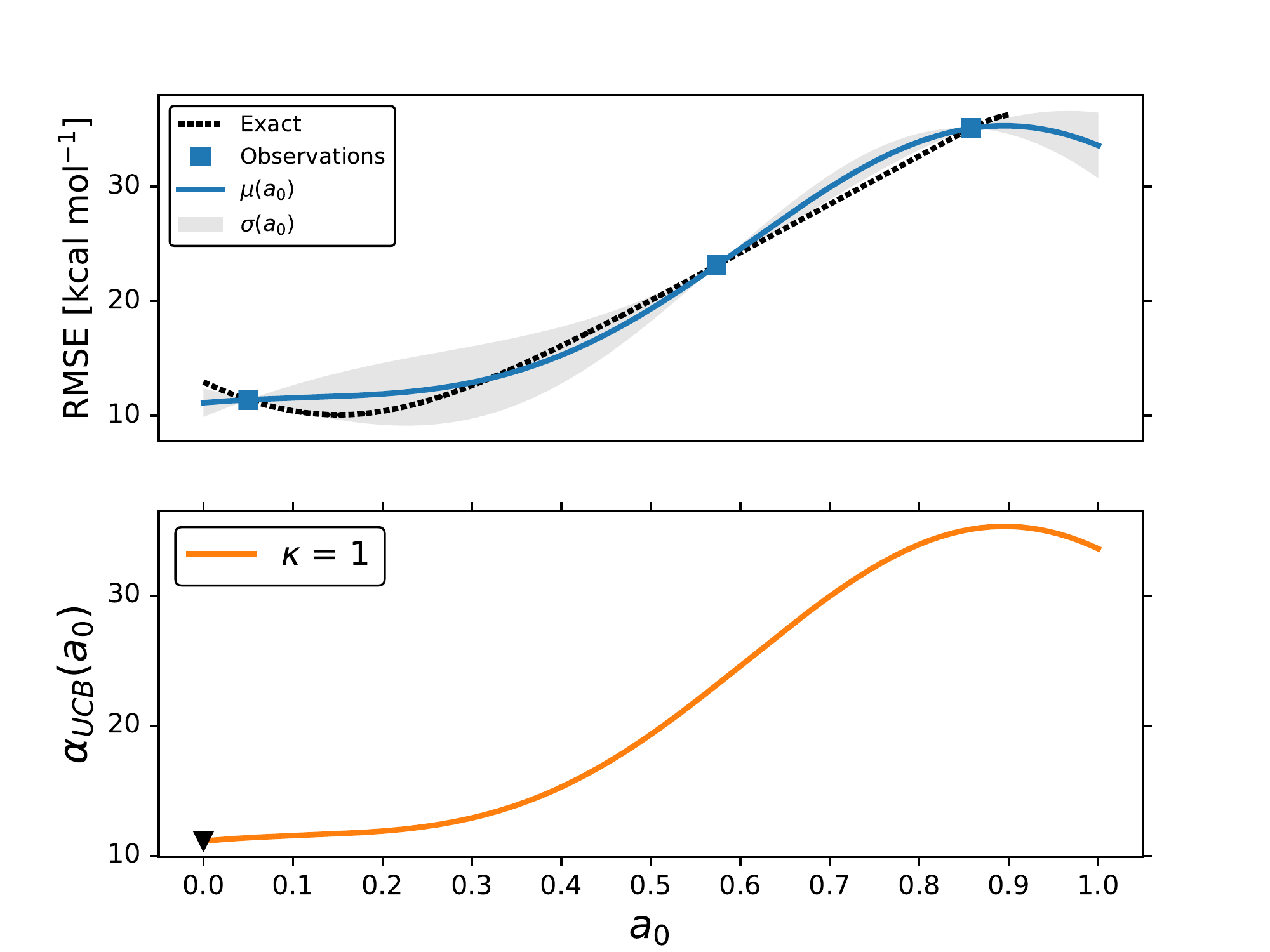}
        \caption{}
        \label{fig:gull}
    \end{subfigure}
    \begin{subfigure}[b]{0.42\textwidth}
        \includegraphics[width=\textwidth]{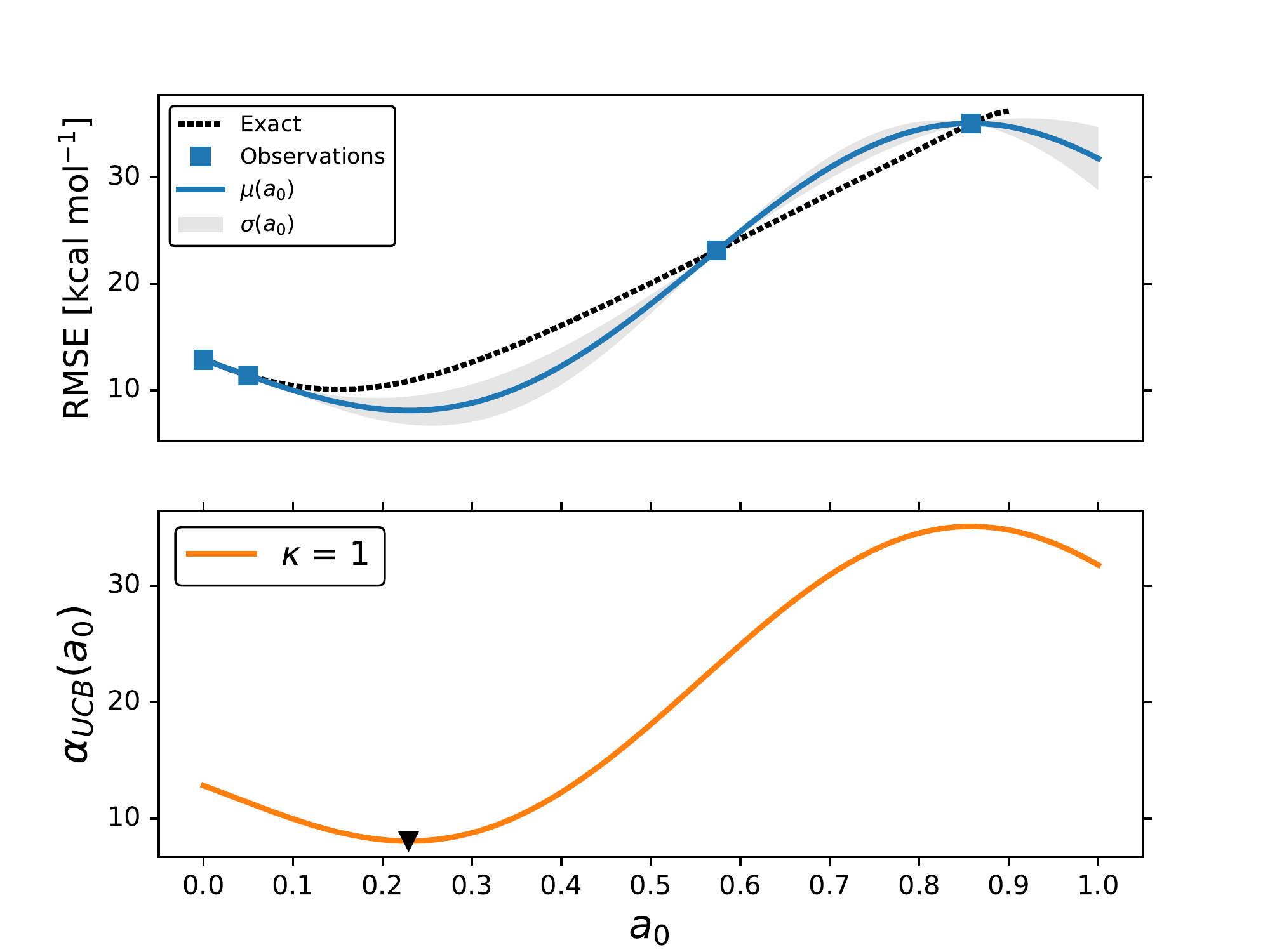}
        \caption{}
        \label{fig:tiger}
    \end{subfigure}
    \begin{subfigure}[b]{0.42\textwidth}
        \includegraphics[width=\textwidth]{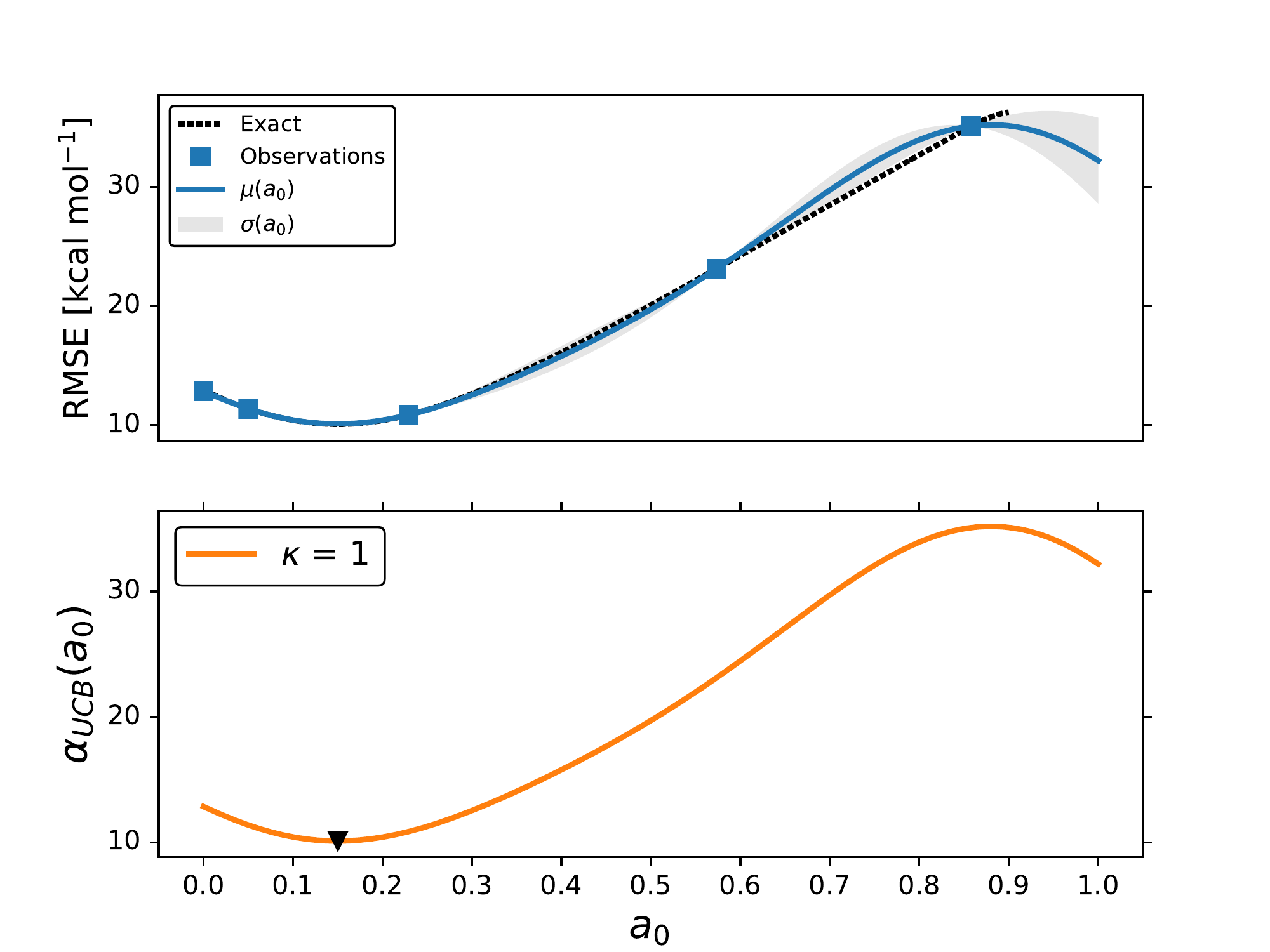}
        \caption{}
        \label{fig:mouse}
    \end{subfigure}
    \begin{subfigure}[b]{0.42\textwidth}
        \includegraphics[width=\textwidth]{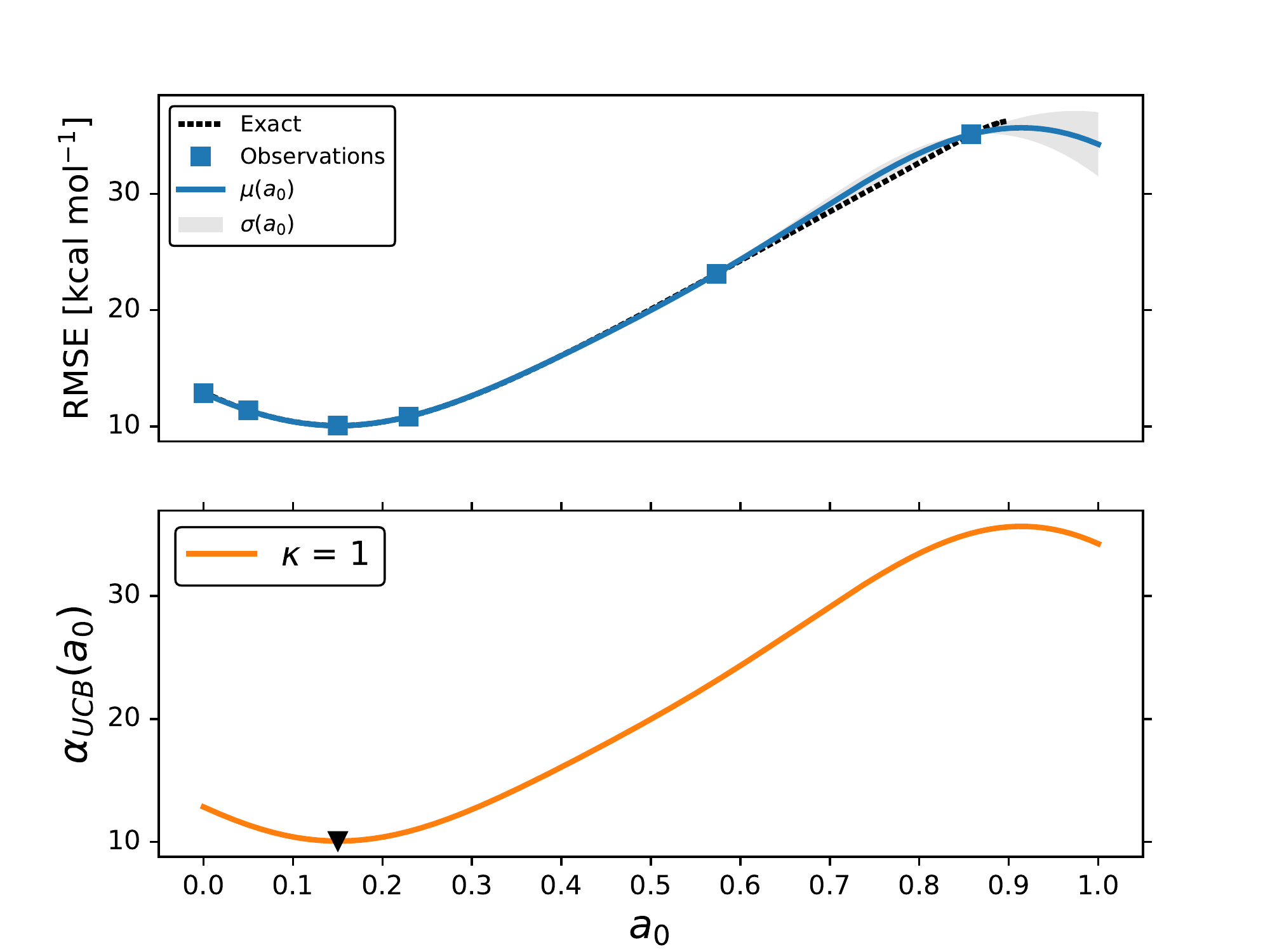}
        \caption{}
        \label{fig:mouse}
    \end{subfigure}    
    \caption{Each figure is an iteration of the BO algorithm, we present present 4 iterations. 
    (Upper panels) The solid-blue curve is the mean of the GP model and shaded area is  the standard deviation of the GP model. The markers are the training data used to construct the GP model. 
    The black-dashed line is the exact form or $f$.  The orange curves in the lower panels are the UCB acquisition function, Eq. \ref{eqn:a_ucb}, with $\kappa =1$.  
    The marker ($\blacktriangledown$) represents the maximum of $\alpha_{UCB}$ which is the proposed point by the BO algorithm.  }\label{fig:BO_PBE0}
\end{figure}

In this section we present how BO algorithm finds the minimum of an error function like Eq. \ref{eqn:loss_dft} for DF models. 
For illustrative porpoises we minimized the single adjustable parameter $a_0$ of the $PBE0$ functional where, $a_X = 1 - a_0$ and $a_C = 1$ \cite{acm_theory0,acm_theory,PBE0_1}.
We only considered the atomization energies in Table \ref{tab:ae_g1}, the Gaussian-1 (G1) database \cite{G1,G1_1,G1_2}.
The first step in the BO algorithm is to gather some data to train a GP model. 
The training data was gathered by randomly sample three values for $a_0$, $x_{1:3} \sim U(0,1)$, and compute $f(a_0, a_X = 1 - a_0, a_C=1 )$. 
We used atin hyper cube sampling (LHS) algorithm \cite{LHS} to sample the initial points when we jointly optimized all three DF parameters. 
The second step is to train a GP model to approximate $f$.
Then we optimized the acquisition function constructed with the trained GP model to select the next query point where $f$ will be evaluated, $f(\mathbf{x}_{N+1})$. The optimization of the acquisition function can be done numerically. 
 Once we evaluate $f$ at $\mathbf{x}_{N+1}$ we update the GP model and carried this procedure sequentially until we converge to the minimum/maximum of $f$.
 Algorithm \ref{alg:BO} and Figure \ref{fig:BO_PBE0} illustrate the BO algorithm to optimize hybrid-DF models.
In Figure \ref{fig:BO_PBE0} we can observe that with only 6 evaluations of $f$, BO found that the minimum of $f$ is for $a_0 = 0.1502$; the $\text{RMSE} = 10.08$ kcal mol$^{-1}$. 
For this calculations we used the UCB acquisition function with $\kappa = 1$. \\

\begin{center}
\begin{minipage}[t]{0.5\textwidth}
\begin{algorithm}[H]
  \caption{Bayesian optimization}\label{alg:BO}
 \textbf{Input:} Acquisition function $\alpha(\cdot)$, black-box function $f(\cdot)$, data set ${\cal D}$.
   \begin{algorithmic}[1]
\For{n = 1,2, $\dots$,} 
\State Optimize the acquisition function,
	\begin{eqnarray*}
	\mathbf{x_{n+1}} = \underset{\mathbf{x}}{\mathrm{arg\;max}} \;\alpha(\mathbf{x},{\cal D})
	\end{eqnarray*}
\State Evaluate $f(\mathbf{x_{n+1}})$.
\State Augment data ${\cal D}_{n+1} = \{ {\cal D}_n, (\mathbf{x_{n+1}},f(\mathbf{x_{n+1}}))\}$.
\State Update model.	
 \EndFor
   \end{algorithmic}
\end{algorithm}
\end{minipage}
\end{center}

\section{Results}
\subsection{Acquisition hyper-parameter}
\begin{figure}[h]
    \centering
        \includegraphics[width=0.5\textwidth]{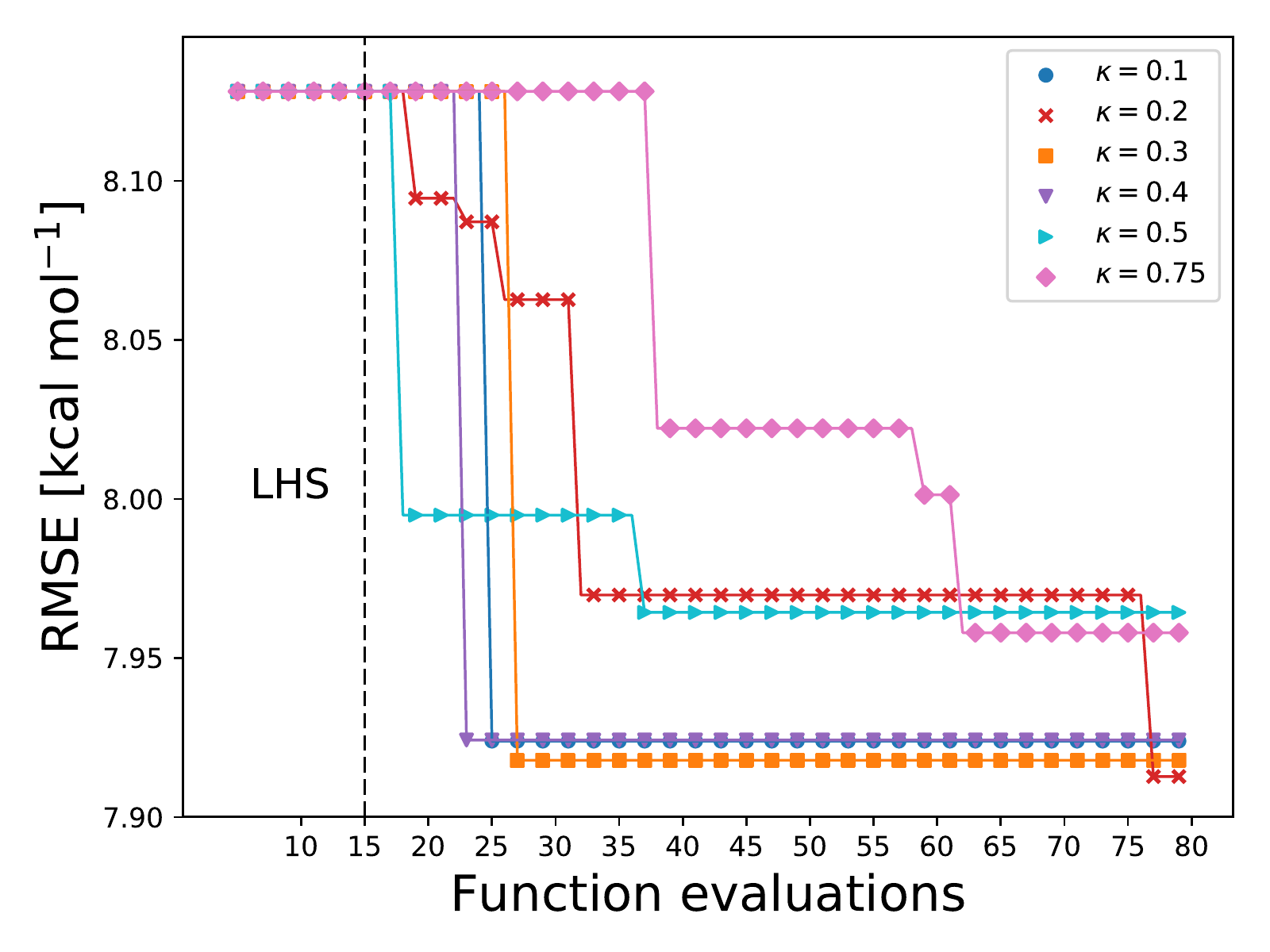}
        \caption{Lowest RMSE at each iteration of the BO algorithm for the optimization of the free parameters of $E_X= B$ and $E_C=LYP$.  We used the UCB acquisition function, Eq. \ref{eqn:a_ucb}, with different values of $\kappa$.
	The RMSE is computed using Eq. \ref{eqn:loss_dft}, where $R_{mi}$ are atomization energies of the G1 molecules, Refs. \cite{G1,G1_1,G1_2}. }
	\label{fig:BO_ucb_k_DFT}
\end{figure}

The hyper-parameter of the acquisition plays a key role for Bayesian optimization. 
For the \emph{upper confidence bound} (UCB) acquisition function, the hyper-parameter, $\kappa$, controls the exploration-exploitation in the BO algorithm.
The convergence of BO towards the optimal value of the DF model's free parameters, $a_0$, $a_X$ and $a_C$, for different values of $\kappa$, Fig. \ref{fig:BO_ucb_k_DFT}.
Furthermore, when $\kappa<0.5$, BO found the lowest RMSE with less than 50 iterations. 

\subsection{Results for the ACM3 functionals}
\begin{center}
\begin{table}[H]
\footnotesize
\caption{The RMSE of various $E_{XC}$ optimized with BO and the UCB acquisition function with $\kappa =0.1$. RMSEs are reported in [kcal mol$^{-1}$] \cite{g1_ae}. The RMSE was computed with respect to atomization energy of G1 molecules \cite{G1,G1_1,G1_2}, Table \ref{tab:ae_g1}.
We sampled 15 initial points with the LHS algorithm.}
\begin{minipage}[b]{0.33\linewidth}
\begin{tabular}{c c | c c c | c }
 $E_X$ & $E_C$ & $a_0$ & $a_X$  & $a_C$ & RMSE  \\ \hline \hline
\multirow{5}{*}{B}&\multirow{5}{*}{LYP}& 0.1055 &  0.8416 &  0.0001 &  7.911  \\ 
 &  &  0.1028 &  0.8484 &  0.0001 &  7.913  \\ 
 &  &  0.1100 &  0.8293 &  0.0201 &  7.912  \\ 
 &  &  0.1112 &  0.8299 &  0.0162 &  7.913  \\ 
 &  &  0.1078 &  0.8374 &  0.0001 &  7.913  \\  \hline
 \multirow{5}{*}{B}&\multirow{5}{*}{P86}& 0.1156 &  0.7842 &  0.0001 &  7.887  \\ 
 &  &  0.1153 &  0.7845 &  0.0001 &  7.887  \\ 
 &  &  0.1178 &  0.7802 &  0.0001 &  7.887  \\ 
 &  &  0.1157 &  0.7837 &  0.0001 &  7.887  \\ 
 &  &  0.1153 &  0.7839 &  0.0001 &  7.887  \\  \hline
 \multirow{5}{*}{B}&\multirow{5}{*}{PW91}& 0.0853 &  0.7864 &  0.0001 &  8.047  \\ 
 &  &  0.0856 &  0.7857 &  0.0001 &  8.047  \\ 
 &  &  0.0852 &  0.7865 &  0.0001 &  8.047  \\ 
 &  &  0.0856 &  0.7858 &  0.0001 &  8.047  \\ 
 &  &  0.0849 &  0.7870 &  0.0001 &  8.047  \\  \hline
\multirow{5}{*}{B}&\multirow{5}{*}{PBE}& 0.0855 &  0.7870 &  0.0001 &  8.047  \\ 
 &  &  0.0855 &  0.7860 &  0.0001 &  8.047  \\ 
 &  &  0.0861 &  0.7847 &  0.0001 &  8.047  \\ 
 &  &  0.0902 &  0.7771 &  0.0001 &  8.048  \\ 
 &  &  0.0860 &  0.7856 &  0.0001 &  8.047  \\ \hline
\multirow{5}{*}{B}&\multirow{5}{*}{VP86}& 0.0830 &  0.7917 &  0.0001 &  8.080  \\ 
 &  &  0.0789 &  0.7981 &  0.0001 &  8.079  \\ 
 &  &  0.0759 &  0.8077 &  0.0001 &  8.081  \\ 
 &  &  0.0786 &  0.7997 &  0.0001 &  8.079  \\ 
 &  &  0.0786 &  0.7994 &  0.0001 &  8.079  \\ \hline
\multirow{5}{*}{B}&\multirow{5}{*}{V5LYP}& 0.0767 &  0.8030 &  0.0001 &  8.079  \\ 
 &  &  0.0774 &  0.8028 &  0.0001 &  8.079  \\ 
 &  &  0.0806 &  0.7945 &  0.0001 &  8.079  \\ 
 &  &  0.0830 &  0.7892 &  0.0001 &  8.080  \\ 
 &  &  0.0790 &  0.7987 &  0.0001 &  8.079  \\ \hline
\multirow{5}{*}{PW91}&\multirow{5}{*}{LYP}& 0.1266 &  0.9247 &  0.1964 &  8.021  \\ 
 &  &  0.1274 &  0.9269 &  0.1869 &  8.021  \\ 
 &  &  0.1173 &  0.9430 &  0.1382 &  8.029  \\ 
 &  &  0.1277 &  0.9242 &  0.1983 &  8.021  \\ 
 &  &  0.1288 &  0.9189 &  0.1877 &  8.022  \\ \hline
\multirow{5}{*}{PW91}&\multirow{5}{*}{P86}& 0.1139 &  0.9218 &  0.0001 &  7.993  \\ 
 &  &  0.1149 &  0.9184 &  0.0001 &  7.993  \\ 
 &  &  0.0989 &  0.9516 &  0.0088 &  8.009  \\ 
 &  &  0.0819 &  1.0000 &  0.0001 &  8.040  \\ 
 &  &  0.1161 &  0.9167 &  0.0001 &  7.993  \\ \hline 
\multirow{5}{*}{PW91}&\multirow{5}{*}{PW91}& 0.0838 &  0.9254 &  0.0034 &  8.100  \\ 
 &  &  0.0935 &  0.9004 &  0.0099 &  8.103  \\ 
 &  &  0.0820 &  0.9293 &  0.0001 &  8.100  \\ 
 &  &  0.0838 &  0.9236 &  0.0001 &  8.100  \\ 
 &  &  0.0831 &  0.9254 &  0.0001 &  8.100  \\ \hline
 \multirow{5}{*}{PW91}&\multirow{5}{*}{PBE}&  0.0826 &  0.9268 &  0.0001 &  8.100  \\ 
 &  &  0.0814 &  0.9304 &  0.0001 &  8.100  \\ 
 &  &  0.0821 &  0.9284 &  0.0001 &  8.100  \\ 
 &  &  0.0826 &  0.9274 &  0.0140 &  8.100  \\ 
 &  &  0.0534 &  1.0000 &  0.0001 &  8.138  \\ 
\end{tabular}
\end{minipage}
\hspace{0.5cm}
\begin{minipage}[b]{0.33\linewidth}
\begin{tabular}{c c | c c c | c }
 $E_X$ & $E_C$ & $a_0$ & $a_X$  & $a_C$ & RMSE  \\ \hline \hline
\multirow{5}{*}{PW91}&\multirow{5}{*}{VP86}&  0.0766 &  0.9406 &  0.0001 &  8.132  \\ 
 &  &  0.0768 &  0.9395 &  0.0001 &  8.132  \\ 
 &  &  0.0790 &  0.9339 &  0.0001 &  8.132  \\ 
 &  &  0.0768 &  0.9399 &  0.0001 &  8.132  \\ 
 &  &  0.0736 &  0.9476 &  0.0001 &  8.132  \\ \hline
\multirow{5}{*}{PW91}&\multirow{5}{*}{V5LYP}&  0.1147 &  0.8844 &  0.2830 &  8.158  \\ 
 &  &  0.0957 &  0.9008 &  0.1392 &  8.123  \\ 
 &  &  0.0931 &  0.9086 &  0.1226 &  8.121  \\ 
 &  &  0.0985 &  0.9028 &  0.1663 &  8.123  \\ 
 &  &  0.0931 &  0.9090 &  0.1174 &  8.121  \\  \hline
\multirow{5}{*}{mPW}&\multirow{5}{*}{LYP}& 0.1151 &  0.8808 &  0.1056 &  7.957  \\ 
 &  &  0.1046 &  0.9064 &  0.0001 &  7.966  \\ 
 &  &  0.1228 &  0.8642 &  0.1283 &  7.959  \\ 
 &  &  0.1142 &  0.8829 &  0.0933 &  7.957  \\ 
 &  &  0.1201 &  0.8720 &  0.0658 &  7.961  \\ \hline
\multirow{5}{*}{mPW}&\multirow{5}{*}{P86}& 0.1130 &  0.8476 &  0.0001 &  7.924  \\ 
 &  &  0.1139 &  0.8457 &  0.0001 &  7.924  \\ 
 &  &  0.1124 &  0.8499 &  0.0001 &  7.924  \\ 
 &  &  0.1136 &  0.8465 &  0.0001 &  7.924  \\ 
 &  &  0.1120 &  0.8497 &  0.0001 &  7.924  \\  \hline
\multirow{5}{*}{mPW}&\multirow{5}{*}{PW91}& 0.0832 &  0.8507 &  0.0001 &  8.060  \\ 
 &  &  0.0834 &  0.8482 &  0.0001 &  8.060  \\ 
 &  &  0.0849 &  0.8451 &  0.0001 &  8.060  \\ 
 &  &  0.0838 &  0.8481 &  0.0001 &  8.060  \\ 
 &  &  0.0829 &  0.8506 &  0.0001 &  8.060  \\ \hline
\multirow{5}{*}{mPW}&\multirow{5}{*}{PBE}&  0.0828 &  0.8503 &  0.0001 &  8.060  \\ 
 &  &  0.0820 &  0.8506 &  0.0001 &  8.060  \\ 
 &  &  0.0731 &  0.8718 &  0.0001 &  8.063  \\ 
 &  &  0.0826 &  0.8510 &  0.0001 &  8.060  \\ 
 &  &  0.0831 &  0.8496 &  0.0001 &  8.060  \\ \hline
\multirow{5}{*}{mPW}&\multirow{5}{*}{VP86}& 0.0772 &  0.8624 &  0.0001 &  8.092  \\ 
 &  &  0.0763 &  0.8647 &  0.0001 &  8.092  \\ 
 &  &  0.0768 &  0.8617 &  0.0001 &  8.092  \\ 
 &  &  0.0768 &  0.8647 &  0.0001 &  8.092  \\ 
 &  &  0.0776 &  0.8617 &  0.0001 &  8.092  \\ \hline 
\multirow{5}{*}{mPW}&\multirow{5}{*}{V5LYP}& 0.0865 &  0.8483 &  0.0706 &  8.090  \\ 
 &  &  0.0820 &  0.8558 &  0.0494 &  8.090  \\ 
 &  &  0.0876 &  0.8458 &  0.0702 &  8.090  \\ 
 &  &  0.0841 &  0.8505 &  0.0472 &  8.090  \\ 
 &  &  0.0864 &  0.8447 &  0.0435 &  8.089  \\ \hline
 \multirow{5}{*}{G96}&\multirow{5}{*}{LYP}& 0.1031 &  0.7865 &  0.0001 &  7.909  \\ 
 &  &  0.1079 &  0.7814 &  0.0001 &  7.912  \\ 
 &  &  0.1024 &  0.7881 &  0.0001 &  7.909  \\ 
 &  &  0.1052 &  0.7826 &  0.0001 &  7.910  \\ 
 &  &  0.1015 &  0.7890 &  0.0001 &  7.909  \\ \hline
 \multirow{5}{*}{G96}&\multirow{5}{*}{ P86}&  0.1146 &  0.7286 &  0.0001 &  7.896  \\ 
 &  &  0.1129 &  0.7325 &  0.0001 &  7.896  \\ 
 &  &  0.1130 &  0.7321 &  0.0001 &  7.896  \\ 
 &  &  0.1131 &  0.7320 &  0.0001 &  7.896  \\ 
 &  &  0.1122 &  0.7334 &  0.0001 &  7.896  \\ 
\end{tabular}
\end{minipage}
\hspace{0.5cm}
\begin{minipage}[b]{0.33\linewidth}
\begin{tabular}{c c | c c c | c } $E_X$ & $E_C$ & $a_0$ & $a_X$  & $a_C$ & RMSE  \\ \hline \hline
\multirow{5}{*}{G96}&\multirow{5}{*}{PW91}& 0.0834 &  0.7327 &  0.0001 &  8.130  \\ 
 &  &  0.0849 &  0.7301 &  0.0001 &  8.130  \\ 
 &  &  0.0854 &  0.7295 &  0.0001 &  8.130  \\ 
 &  &  0.0839 &  0.7318 &  0.0001 &  8.130  \\ 
 &  &  0.0840 &  0.7320 &  0.0001 &  8.130  \\ \hline 
\multirow{5}{*}{G96}&\multirow{5}{*}{PBE}& 0.0838 &  0.7327 &  0.0001 &  8.130  \\ 
 &  &  0.0844 &  0.7319 &  0.0010 &  8.130  \\ 
 &  &  0.0841 &  0.7312 &  0.0001 &  8.130  \\ 
 &  &  0.0852 &  0.7283 &  0.0001 &  8.130  \\ 
 &  &  0.0836 &  0.7325 &  0.0001 &  8.130  \\ \hline 
\multirow{5}{*}{G96}&\multirow{5}{*}{VP86}& 0.0757 &  0.7459 &  0.0004 &  8.168  \\ 
 &  &  0.0782 &  0.7425 &  0.0001 &  8.168  \\ 
 &  &  0.0783 &  0.7426 &  0.0001 &  8.168  \\ 
 &  &  0.0773 &  0.7435 &  0.0001 &  8.168  \\ 
 &  &  0.0777 &  0.7463 &  0.0001 &  8.169  \\ \hline
 \multirow{5}{*}{G96}&\multirow{5}{*}{V5LYP}& 0.0803 &  0.7378 &  0.0001 &  8.168  \\ 
 &  &  0.0751 &  0.7471 &  0.0001 &  8.168  \\ 
 &  &  0.0682 &  0.7563 &  0.0001 &  8.174  \\ 
 &  &  0.0784 &  0.7415 &  0.0007 &  8.168  \\ 
 &  &  0.0010 &  0.5006 &  1.0000 &  8.168  \\ \hline
 \multirow{5}{*}{PBE}&\multirow{5}{*}{LYP}& 0.1145 &  0.9536 &  0.0001 &  8.033  \\ 
 &  &  0.1287 &  0.9136 &  0.1413 &  8.014  \\ 
 &  &  0.1311 &  0.9075 &  0.1445 &  8.014  \\ 
 &  &  0.1324 &  0.9040 &  0.1497 &  8.014  \\ 
 &  &  0.1333 &  0.9024 &  0.1466 &  8.014  \\ \hline 
 \multirow{5}{*}{PBE}&\multirow{5}{*}{P86}& 0.1247 &  0.8845 &  0.0001 &  7.984  \\ 
 &  &  0.1148 &  0.9057 &  0.0001 &  7.988  \\ 
 &  &  0.1328 &  0.8676 &  0.0001 &  7.986  \\ 
 &  &  0.1241 &  0.8868 &  0.0001 &  7.984  \\ 
 &  &  0.1251 &  0.8839 &  0.0001 &  7.984  \\  \hline 
 \multirow{5}{*}{PBE}&\multirow{5}{*}{PW91}& 0.0895 &  0.8997 &  0.0001 &  8.072  \\ 
 &  &  0.0898 &  0.8986 &  0.0001 &  8.072  \\ 
 &  &  0.0921 &  0.8937 &  0.0001 &  8.071  \\ 
 &  &  0.0994 &  0.8752 &  0.0001 &  8.074  \\ 
 &  &  0.0921 &  0.8935 &  0.0001 &  8.071  \\  \hline 
 \multirow{5}{*}{PBE}&\multirow{5}{*}{PBE}&  0.0924 &  0.8920 &  0.0001 &  8.071  \\ 
 &  &  0.0910 &  0.8986 &  0.0001 &  8.072  \\ 
 &  &  0.1028 &  0.8708 &  0.0292 &  8.079  \\ 
 &  &  0.0923 &  0.8933 &  0.0001 &  8.071  \\ 
 &  &  0.0913 &  0.8957 &  0.0001 &  8.071  \\ \hline
 \multirow{5}{*}{PBE}&\multirow{5}{*}{VP86}&  0.0861 &  0.9079 &  0.0001 &  8.098  \\ 
 &  &  0.0838 &  0.9128 &  0.0001 &  8.098  \\ 
 &  &  0.0864 &  0.9068 &  0.0001 &  8.098  \\ 
 &  &  0.0909 &  0.8927 &  0.0001 &  8.100  \\ 
 &  &  0.0844 &  0.9109 &  0.0001 &  8.098  \\ \hline
\multirow{5}{*}{PBE}&\multirow{5}{*}{V5LYP}& 0.0949 &  0.8918 &  0.0721 &  8.097  \\ 
 &  &  0.0955 &  0.8891 &  0.0594 &  8.097  \\ 
 &  &  0.0881 &  0.9038 &  0.0282 &  8.097  \\ 
 &  &  0.0858 &  0.9073 &  0.0001 &  8.098  \\ 
 &  &  0.1013 &  0.8760 &  0.0272 &  8.103  \\
\end{tabular}
\end{minipage}
\label{tab:acm3_exc_ucb}
\end{table}
\end{center}


\begin{table}
\footnotesize
\caption{The RMSE of various $E_{XC}$ optimized with BO and the UCB acquisition function with $\kappa =0.1$. RMSEs and MAEs are reported in [$\AA$] \cite{g1_ae}. The RMSE was computed with respect to experimental bond lengths of G1 diatomic molecules \cite{G1,G1_1,G1_2}, Table \ref{tab:bl_g1}.
We sampled 15 initial points with the LHS algorithm.}
 \begin{minipage}[b]{0.4\linewidth}
\begin{tabular}{ c | c  c c | c  c } $E_X-E_C$ & $a_0$ & $a_X$  & $a_C$ & RMSE  & MAE \\ \hline \hline
\multirow{3}{*}{ B-LYP }  &  0.1467 &  0.6323 &  0.0010 &  0.00704 &  0.00419 \\ 
  &   0.1505 &  0.6682 &  0.0094 &  0.00709 &  0.00442 \\ 
  &   0.1450 &  0.6266 &  0.0010 &  0.00705 &  0.00415 \\ \hline
\multirow{3}{*}{ B-P86 }  &  0.1933 &  0.6930 &  0.0010 &  0.01007 &  0.00518 \\ 
  &   0.2039 &  0.7117 &  0.0010 &  0.01006 &  0.00537 \\ 
  &   0.1981 &  0.7104 &  0.0010 &  0.01006 &  0.00529 \\  \hline
\multirow{3}{*}{ B-PW91 }  &  0.1941 &  0.6997 &  0.0010 &  0.01033 &  0.00535 \\ 
  &   0.1936 &  0.7024 &  0.0010 &  0.01033 &  0.00535 \\ 
  &   0.1977 &  0.6999 &  0.0010 &  0.01034 &  0.00540 \\  \hline
\multirow{3}{*}{ B-PBE }  &  0.2361 &  0.9537 &  1.0000 &  0.01138 &  0.00769 \\ 
  &   0.1939 &  0.7024 &  0.0010 &  0.01034 &  0.00535 \\ 
  &   0.1939 &  0.7015 &  0.0010 &  0.01033 &  0.00535 \\  \hline
\multirow{3}{*}{ B-VP86 }  &  0.1895 &  0.6912 &  0.0010 &  0.01027 &  0.00529 \\ 
  &   0.1856 &  0.6574 &  0.0010 &  0.01028 &  0.00517 \\ 
  &   0.1921 &  0.6964 &  0.0010 &  0.01027 &  0.00533 \\  \hline
\multirow{3}{*}{ B-V5LYP }  &  0.2464 &  0.8518 &  0.9277 &  0.00984 &  0.00537 \\ 
  &   0.2426 &  0.9088 &  1.0000 &  0.00971 &  0.00497 \\ 
  &   0.2531 &  0.9272 &  1.0000 &  0.00971 &  0.00513 \\  \hline
\multirow{3}{*}{ PW91-LYP }  &  0.1388 &  0.6970 &  0.0010 &  0.00537 &  0.00407 \\ 
  &   0.1546 &  0.7163 &  1.0000 &  0.00548 &  0.00415 \\ 
  &   0.1383 &  0.6920 &  0.0010 &  0.00537 &  0.00405 \\  \hline
\multirow{3}{*}{ PW91-P86 }  &  0.2158 &  0.9862 &  0.1850 &  0.00827 &  0.00555 \\ 
  &   0.1909 &  0.8101 &  0.0013 &  0.00750 &  0.00486 \\ 
  &   0.1929 &  0.8012 &  0.0010 &  0.00749 &  0.00479 \\  \hline
\multirow{3}{*}{ PW91-PW91 }  &  0.1878 &  0.8020 &  0.0010 &  0.00773 &  0.00484 \\ 
  &   0.1864 &  0.7979 &  0.0010 &  0.00773 &  0.00483 \\ 
  &   0.1912 &  0.7953 &  0.0010 &  0.00773 &  0.00481 \\  \hline
\multirow{3}{*}{ PW91-PBE }  &  0.1602 &  0.9386 &  0.0010 &  0.00858 &  0.00623 \\ 
  &   0.1841 &  0.7910 &  0.0010 &  0.00774 &  0.00481 \\ 
  &   0.1878 &  0.8011 &  0.0010 &  0.00773 &  0.00484 \\  \hline
\multirow{3}{*}{ PW91-VP86 }  &  0.1419 &  0.6777 &  0.0010 &  0.00821 &  0.00428 \\ 
  &   0.1836 &  0.8024 &  0.0135 &  0.00774 &  0.00484 \\ 
  &   0.1833 &  0.7927 &  0.0010 &  0.00769 &  0.00481 \\  \hline
\multirow{3}{*}{ PW91-V5LYP }  &  0.2321 &  0.9910 &  1.0000 &  0.00673 &  0.00488 \\ 
  &   0.2231 &  1.0000 &  1.0000 &  0.00673 &  0.00492 \\ 
  &   0.2143 &  0.9576 &  0.8591 &  0.00685 &  0.00484 \\  \hline
\multirow{3}{*}{ mPW-LYP }  &  0.1434 &  0.6679 &  0.0010 &  0.00617 &  0.00408 \\ 
  &   0.1409 &  0.6524 &  0.0010 &  0.00616 &  0.00403 \\ 
  &   0.1558 &  0.7262 &  0.0067 &  0.00630 &  0.00444 \\  \hline
\multirow{3}{*}{ mPW-P86 }  &  0.1949 &  0.7492 &  0.0010 &  0.00883 &  0.00504 \\ 
  &   0.1972 &  0.7436 &  0.0010 &  0.00883 &  0.00501 \\ 
  &   0.1934 &  0.7455 &  0.0010 &  0.00883 &  0.00501 \\  \hline
\multirow{3}{*}{ mPW-PW91 }  &  0.1930 &  0.7491 &  0.0010 &  0.00910 &  0.00512 \\ 
  &   0.1938 &  0.7575 &  0.0010 &  0.00909 &  0.00514 \\ 
  &   0.1884 &  0.7404 &  0.0010 &  0.00911 &  0.00505 \\  \hline
  \end{tabular}
 \end{minipage}
 \hspace{0.5cm}
 \begin{minipage}[b]{0.4\linewidth}
 \begin{tabular}{c | c  c  c | c  c } $E_X-E_C$ & $a_0$ & $a_X$  & $a_C$ & RMSE  & MAE \\ \hline \hline
\multirow{3}{*}{ mPW-PBE }  &  0.1961 &  0.7609 &  0.0010 &  0.00909 &  0.00516 \\ 
  &   0.1967 &  0.7605 &  0.0010 &  0.00909 &  0.00517 \\ 
  &   0.1884 &  0.7271 &  0.0281 &  0.00933 &  0.00495 \\  \hline
\multirow{3}{*}{ mPW-VP86 }  &  0.1881 &  0.7575 &  0.0010 &  0.00904 &  0.00514 \\ 
  &   0.2194 &  0.8163 &  0.0010 &  0.00926 &  0.00585 \\ 
  &   0.1905 &  0.7486 &  0.0010 &  0.00904 &  0.00509 \\  \hline
\multirow{3}{*}{ mPW-V5LYP }  &  0.2957 &  1.0000 &  1.0000 &  0.00871 &  0.00612 \\ 
  &   0.2355 &  0.9268 &  0.8757 &  0.00830 &  0.00495 \\ 
  &   0.2483 &  1.0000 &  1.0000 &  0.00821 &  0.00518 \\  \hline
\multirow{3}{*}{ G96-LYP }  &  0.1283 &  0.5508 &  0.0010 &  0.00930 &  0.00507 \\ 
  &   0.1328 &  0.5620 &  0.0010 &  0.00931 &  0.00518 \\ 
  &   0.1291 &  0.5712 &  0.0010 &  0.00931 &  0.00505 \\  \hline
\multirow{3}{*}{ G96-P86 }  &  0.1821 &  0.5883 &  0.0010 &  0.01280 &  0.00619 \\ 
  &   0.1835 &  0.5736 &  0.0010 &  0.01279 &  0.00630 \\ 
  &   0.1793 &  0.5432 &  0.0010 &  0.01280 &  0.00642 \\  \hline
\multirow{3}{*}{ G96-PW91 }  &  0.1703 &  0.5501 &  0.0010 &  0.01306 &  0.00639 \\ 
  &   0.1755 &  0.5588 &  0.0010 &  0.01308 &  0.00641 \\ 
  &   0.1829 &  0.5915 &  0.0010 &  0.01307 &  0.00632 \\  \hline
\multirow{3}{*}{ G96-PBE }  &  0.1744 &  0.5642 &  0.0010 &  0.01308 &  0.00637 \\ 
  &   0.1715 &  0.5594 &  0.0010 &  0.01307 &  0.00635 \\ 
  &   0.1842 &  0.5832 &  0.0010 &  0.01307 &  0.00637 \\  \hline
\multirow{3}{*}{ G96-VP86 }  &  0.1690 &  0.5519 &  0.0010 &  0.01299 &  0.00639 \\ 
  &   0.1693 &  0.5637 &  0.0010 &  0.01298 &  0.00631 \\ 
  &   0.1677 &  0.5577 &  0.0010 &  0.01299 &  0.00633 \\  \hline
\multirow{3}{*}{ G96-V5LYP }  &  0.1705 &  0.5668 &  0.0010 &  0.01298 &  0.00630 \\ 
  &   0.1828 &  0.5700 &  0.0010 &  0.01298 &  0.00639 \\ 
  &   0.1644 &  0.5533 &  0.0010 &  0.01298 &  0.00631 \\  \hline
\multirow{3}{*}{ PBE-LYP }  &  0.1586 &  0.7236 &  0.0168 &  0.00506 &  0.00385 \\ 
  &   0.1559 &  0.6825 &  0.0592 &  0.00508 &  0.00373 \\ 
  &   0.1440 &  0.6796 &  0.0010 &  0.00498 &  0.00372 \\  \hline
\multirow{3}{*}{ PBE-P86 }  &  0.2059 &  0.8107 &  0.0010 &  0.00698 &  0.00467 \\ 
  &   0.1997 &  0.7158 &  0.0015 &  0.00713 &  0.00427 \\ 
  &   0.2019 &  0.7843 &  0.0010 &  0.00698 &  0.00454 \\  \hline
\multirow{3}{*}{ PBE-PW91 }  &  0.2128 &  0.8170 &  0.0512 &  0.00750 &  0.00479 \\ 
  &   0.2013 &  0.8035 &  0.0010 &  0.00720 &  0.00468 \\ 
  &   0.2115 &  0.9515 &  0.2803 &  0.00861 &  0.00514 \\  \hline
\multirow{3}{*}{ PBE-PBE }  &  0.1979 &  0.7957 &  0.0010 &  0.00721 &  0.00467 \\ 
  &   0.1969 &  0.8067 &  0.0010 &  0.00721 &  0.00471 \\ 
  &   0.1974 &  0.7901 &  0.0010 &  0.00721 &  0.00463 \\  \hline
\multirow{3}{*}{ PBE-VP86 }  &  0.1910 &  0.7771 &  0.0010 &  0.00718 &  0.00455 \\ 
  &   0.1891 &  0.7847 &  0.0010 &  0.00718 &  0.00461 \\ 
  &   0.1932 &  0.8016 &  0.0010 &  0.00717 &  0.00469 \\  \hline
\multirow{3}{*}{ PBE-V5LYP }  &  0.2253 &  0.9294 &  0.9106 &  0.00625 &  0.00433 \\ 
  &   0.2267 &  0.9262 &  0.7011 &  0.00634 &  0.00444 \\ 
  &   0.2368 &  0.9307 &  0.8924 &  0.00625 &  0.00437 \\  \hline
  \end{tabular}
   \end{minipage}
\label{tab:acm3_exc_ucb_bl}
\end{table}

\begin{table}
\centering
\caption{The predicted RMSE and MAE with the PBE0 and B3LYP functionals computed with respect to the atomization energies of G1 molecules \cite{G1,G1_1,G1_2}, Table \ref{tab:ae_g1}. The molecular geometries were optimized with each basis set.}
\begin{ruledtabular}
\begin{tabular}{  c l | c  c } 
 DFT & basis set & RMSE $[$kcal mol$^{-1}]$  & MAE $[$kcal mol$^{-1}]$  \\ \hline \hline
 PBE0 & 6-31G($d$) & 11.2 & 7.461 \\
 PBE0 & 6-311G($d,p$) & 9.22 & 5.967 \\
 PBE0 & 6-311++G($d,p$) &8.668 & 5.702 \\ 
 PBE0 & 6-311++G($df,pd$) & 8.045 & 4.776 \\ 
 PBE0 & 6-311++G($3df,3pd$) & 7.378 & 3.845 \\  \hline
 B3LYP & 6-31G($d$) & 9.4 & 5.754 \\ 
 B3LYP & 6-311G($d,p$) & 7.522 & 4.237 \\ 
 B3LYP & 6-311++G($d,p$) & 7.156 & 4.162 \\ 
 B3LYP & 6-311++G($df,pd$) & 6.574 & 3.291 \\
 B3LYP & 6-311++G($3df,3pd$) & 6.35 & 3.067 
 \end{tabular}
 \end{ruledtabular}
\label{tab:pbe0_b3lyp_ae_diff_basisset}
\end{table}

\begin{table}
\centering
\caption{The predicted RMSE and MAE of the combinations of two $E_{XC}$, PBE-PBE and B-LYP for different basis sets. The three free coefficients, $a_0$, $a_X$ and $a_C$, of both $E_{XC}$ were optimized with BO, \ref{tab:ae_g1}. 
The optimization of $a_0$, $a_X$ and $a_C$ was carried with respect to the RMSE of the atomization energies of G1 molecules \cite{G1,G1_1,G1_2}, Table \ref{tab:ae_g1}. 
The molecular geometries used to compute the RMSE and MAE  were optimized with MP2/6-31G($d$), Table \ref{tab:acm3_exc_ucb}.
}
\begin{ruledtabular}
\begin{tabular}{ c  l |  c l | c  c } 
\multicolumn{2}{c|}{Molecular geometries} & \multicolumn{2}{c|}{method} &  &\\
method & basis set & $E_X-E_C$ & basis set &  RMSE $[$kcal mol$^{-1}]$  & MAE $[$kcal mol$^{-1}]$ \\ \hline \hline
\multirow{3}{*}{MP2} & \multirow{3}{5em}{6-31G($d$)} & PBE-PBE\footnotemark[1] & 6-31G($d,p$) & 8.079 & 5.132 \\
& & PBE-PBE\footnotemark[1]  & 6-311G($d,p$) & 6.847& 4.194 \\
& & PBE-PBE\footnotemark[1]  & 6-311++G($d,p$) & 6.65& 4.203 \\
 \hline
\multirow{3}{*}{MP2} & \multirow{3}{5em}{6-31G($d$)} & B-LYP\footnotemark[2]  & 6-31G($d,p$) & 7.921& 4.957 \\
 & & B-LYP\footnotemark[2]  & 6-311G($d,p$) & 6.937 & 4.452 \\
&  & B-LYP\footnotemark[2]  & 6-311++G($d,p$) & 6.881 & 4.618
\end{tabular}
\end{ruledtabular}
\footnotetext[1]{ $a_0$ = 0.110,  $a_X$ =  0.8708 , $a_C$ = 0.0292, optimized with BO and with MP2/6-31G($d$)  geometries; Table \ref{tab:acm3_exc_ucb}.}
\footnotetext[2]{$a_0$ =  0.11,  $a_X$ = 0.8293 , $a_C$ = 0.0201, optimized with BO and with MP2/6-31G($d$)  geometries; Table \ref{tab:acm3_exc_ucb}.}
\label{tab:pbepbe_blyp_ae_diff_basisset}
\end{table}

\begin{table}
\centering
\caption{The RMSE and MAE of  PBE-PBE and B-LYP with optimized free parameters for different basis sets. 
We used the optimized values of $a_0$, $a_X$ and $a_C$ reported in Table \ref{tab:acm3_exc_ucb}.
Each molecular geometry was optimized with each specific basis set. 
}
\begin{ruledtabular}
\begin{tabular}{ c  l |  c l | c  c } 
\multicolumn{2}{c|}{Molecular geometries} & \multicolumn{2}{c|}{method} &  &\\
$E_X-E_C$ & basis set & $E_X-E_C$ & basis set & RMSE $[$kcal mol$^{-1}]$ & MAE $[$kcal mol$^{-1}]$  \\ \hline \hline
PBE-PBE$^{a}$ & 6-311G($d$) & PBE-PBE$^{a}$ & 6-311G($d,p$) & 6.91 & 4.392 \\
PBE-PBE$^{a}$ & 6-311++G($d,p$) & PBE-PBE$^{a}$ & 6-311++G($d,p$) & 6.67 & 4.279 \\ \hline
B-LYP$^{b}$ & 6-311G($d,p$)  & B-LYP$^{b}$ & 6-311G($d,p$) & 6.998 & 4.632  \\
B-LYP$^{b}$& 6-311++G($d,p$)  & B-LYP$^{b}$& 6-311++G($d,p$) & 6.909 & 4.178
\end{tabular}
\end{ruledtabular}
\footnotetext[1]{ $a_0$ = 0.110,  $a_X$ =  0.8708 , $a_C$ = 0.0292, optimized with BO and with MP2/6-31G($d$)  geometries; Table \ref{tab:acm3_exc_ucb}.}
\footnotetext[2]{$a_0$ =  0.11,  $a_X$ = 0.8293 , $a_C$ = 0.0201, optimized with BO and with MP2/6-31G($d$)  geometries; Table \ref{tab:acm3_exc_ucb}.}
\label{tab:pbepbe_blyp_ae_diff_basisset_2}
\end{table}


\begin{table}
\caption{The predicted RMSE and MAE of the  PBE-PBE and B-LYP functionals with two different basis sets. 
The reported values of $a_0$, $a_X$ and $a_C$  are the minimizer of the RMSE of the atomization energies of G1 molecules \cite{G1,G1_1,G1_2} found by BO at each optimization. 
The values of, $a_0$, $a_X$ and $a_C$, for both XC functionals were optimized with BO and the UCB acquisition function with $\kappa = 0.1$. For each optimization, we sampled 15 initial points with LHS.
The molecular geometries used for computing the RMSE and MAE were optimized with two different basis sets.
   }
\begin{ruledtabular}
\begin{tabular}{ c | c  c c | c c } 
$E_X-E_C$ & $a_0$ & $a_X$  & $a_C$ & RMSE$[$kcal mol$^{-1}]$ & MAE $[$kcal mol$^{-1}]$   \\ \hline \hline
 \multirow{3}{*}{ PBE-PBE\footnotemark[1]}  &  0.1645 &  0.7886 & 0.3352 & 6.852 & 3.64 \\ 
 & 0.1173 & 0.8742 & 0.1967 & 6.763 & 3.853  \\ 
 & 0.089 &  0.9360 &  0.0698 & 6.816 & 4.108 \\ 
 & 0.1601 &  0.7872 & 0.3761 & 6.861 & 3.649 \\  \hline  
  \multirow{3}{*}{ PBE-PBE\footnotemark[2]}  &  0.1293 &  0.8855 & 0.3537 & 6.326 & 3.405 \\ 
 & 0.1064 & 0.9336 & 0.3633 &  6.353 & 3.415  \\ 
 & 0.1196 & 0.9075 & 0.3410 & 6.33 & 3.425\\  
 & 0.1550 & 0.8250 &  0.3924 &  6.354 & 3.429 \\   \hline
   \multirow{3}{*}{ B-LYP\footnotemark[1]}  &  0.1841 & 0.6820 & 0.6649 & 6.736 & 3.415\\ 
  &   0.1500 &  0.7658 &  0.3748 & 6.669  & 3.711 \\ 
  &   0.1399 &  0.7817 &  0.4681 &  6.691  & 3.577 \\  \hline   
\multirow{3}{*}{ B-LYP\footnotemark[2]}  &  0.1924 & 0.6962 & 0.7080 & 6.263 & 3.094   \\  
  & 0.1506 & 0.7784 & 0.6111 &6.288 & 3.261  \\ 
  & 0.2214 & 0.6378 &  0.8547 & 6.306 & 3.011 \\
  & 0.2213 & 0.6354 & 0.9662 & 6.359 & 2.914 \\ 
  \end{tabular}
\end{ruledtabular}
\footnotetext[1]{6-311G($d,p$)}
\footnotetext[2]{6-311++G($df,pd$)}
\label{tab:pbepbe_blyp_ae_bo_diff_basisset}
\end{table}

\begin{table}
\caption{The predicted RMSE and MAE of the  PBE-PBE and B-LYP functionals with two different basis sets. 
The reported values of $a_0$, $a_X$ and $a_C$  are the minimizer of the MAE of the atomization energies of G1 molecules \cite{G1,G1_1,G1_2} found by BO at each optimization. 
The values of, $a_0$, $a_X$ and $a_C$, for both XC functionals were optimized with BO and the UCB acquisition function with $\kappa = 0.1$. For each optimization, we sampled 15 initial points with LHS.
The molecular geometries used for computing the RMSE and MAE were optimized with 6-311G($d,p$).
   }
\begin{ruledtabular}
\begin{tabular}{ c | c  c c | c c } 
$E_X-E_C$ & $a_0$ & $a_X$  & $a_C$ & RMSE$[$kcal mol$^{-1}]$ & MAE $[$kcal mol$^{-1}]$   \\ \hline \hline
 \multirow{3}{*}{ PBE-PBE}  & 0.1231 & 0.8977 & 0.4960 & 7.015 & 3.398 \\ 
 & 0.1218 & 0.8892 & 0.4490 & 6.911 & 3.403 \\ 
 & 0.1243 & 0.8982 & 0.4423 & 6.971 & 3.410 \\  \hline  
  \multirow{3}{*}{B-LYP}  &  0.1660 & 0.7216 & 0.7705 &  6.813 & 3.139  \\ 
 & 0.1663 & 0.7233 & 0.7827 & 6.838 & 3.134 \\  
 & 0.1652 & 0.7289 & 0.7387 & 6.804 & 3.152 \\  
  \end{tabular}
\end{ruledtabular}
\label{tab:pbepbe_blyp_ae_bo_mae_diff_basisset}
\end{table}

\subsection{Integer-valued variables in BO}

\begin{table}[H]
\centering
\footnotesize
\caption{The lowest RMSE found with BO with the UCB acquisition function with different values of $\kappa$. BO jointly optimizes the functional form,  $E_X$ and $E_C$, and the DF model's free parameters, $a_0, a_X$ and $a_C$.  
The RMSE was computed with respect to atomization energy of G1 molecules \cite{G1,G1_1,G1_2}, Table \ref{tab:ae_g1}.
For each optimization we sampled 30 initial points with the LHS algorithm, including a random XC functional.}
\begin{minipage}[b]{0.4\linewidth}
\caption*{{\bf With floor function} and $\kappa = 0.005$}
\begin{tabular}{c c | c c c | c } 
$E_X$ & $E_C$ & $a_0$ & $a_X$  & $a_C$ & \multirow{2}{*}{RMSE} \\
& & & & & $[$kcal mol$^{-1}]$ \\ \hline \hline
B &  LYP &  0.1558  &   0.7269  &   0.2588  &   7.991 \\
mPW &  P86 &  0.1436  &   0.7761  &   0.0010  &   7.965 \\
PBE &  P86 &  0.1186  &   0.9029  &   0.0010  &   7.989 \\
B &  P86 &  0.0632  &   0.8805  &   0.0010  &   7.998 \\
mPW &  P86 &  0.1019  &   0.8743  &   0.0010  &   7.930 \\
\end{tabular}
\end{minipage}
\hspace{0.5cm}
\begin{minipage}[b]{0.4\linewidth}
\caption*{{\bf Without floor function} and $\kappa = 0.005$}
\begin{tabular}{c c | c c c | c } 
$E_X$ & $E_C$ & $a_0$ & $a_X$  & $a_C$ & \multirow{2}{*}{RMSE} \\
& & & & & $[$kcal mol$^{-1}]$ \\ \hline \hline
B &  LYP &  0.0980  &   0.8181  &   0.0010  &   8.093 \\
mPW &  P86 &  0.1551  &   0.7747  &   0.0010  &   8.010 \\
B &  LYP &  0.1496  &   0.7554  &   0.0456  &   7.970 \\
B &  LYP &  0.1347  &   0.7959  &   0.0010  &   7.956 \\
B &  LYP &  0.1606  &   0.7465  &   0.0010  &   8.044 \\
\end{tabular}
\end{minipage}
\begin{minipage}[b]{0.4\linewidth}
\vspace{1.2cm}
\caption*{{\bf With floor function} and $\kappa = 0.05$}
\begin{tabular}{c c | c c c | c } 
$E_X$ & $E_C$ & $a_0$ & $a_X$  & $a_C$ & \multirow{2}{*}{RMSE} \\
& & & & & $[$kcal mol$^{-1}]$ \\ \hline \hline
B &  P86 &  0.1551  &   0.7036  &   0.0010  &   7.950 \\
B &  LYP &  0.1108  &   0.8276  &   0.0010  &   7.917 \\
B &  P86 &  0.1212  &   0.7613  &   0.0010  &   7.906 \\
G96 &  LYP &  0.1071  &   0.7774  &   0.0010  &   7.913 \\
B &  PBE &  0.0941  &   0.7708  &   0.0010  &   8.051 \\
\end{tabular}
\end{minipage}
\begin{minipage}[b]{0.4\linewidth}
\caption*{{\bf Without floor function} and $\kappa = 0.05$}
\begin{tabular}{c c | c c c | c } 
$E_X$ & $E_C$ & $a_0$ & $a_X$  & $a_C$ & \multirow{2}{*}{RMSE} \\
& & & & & $[$kcal mol$^{-1}]$ \\ \hline \hline
B &  LYP &  0.1283  &   0.7678  &   0.0010  &   8.047 \\
PBE &  LYP &  0.1135  &   0.9667  &   0.0559  &   8.043 \\
B &  LYP &  0.1017  &   0.8479  &   0.0010  &   7.914 \\
PBE &  P86 &  0.1439  &   0.8371  &   0.0433  &   8.040 \\
B &  P86 &  0.1192  &   0.7987  &   0.0010  &   7.944 \\
\end{tabular}
\end{minipage}
\label{tab:sm_exc_dft_k}
\end{table}

The selection of the exchange, $E_X$, and correlation, $E_C$, functionals determines the accuracy in the prediction of physical properties with DF models.
In this work we present that BO can optimize the DF parameters and select the most optimal forms for the exchange and correlation functionals.
For this procedure we considered the same loss function as before, Eq. \ref{eqn:loss_dft}, and include the possibility to select different $E_X$ and $E_C$,
\begin{eqnarray}
{\cal L} = f(E_X,E_C, a_0,a_X,a_C).
\end{eqnarray}

\begin{figure}[H]
\centering
\includegraphics[width=0.45\textwidth]{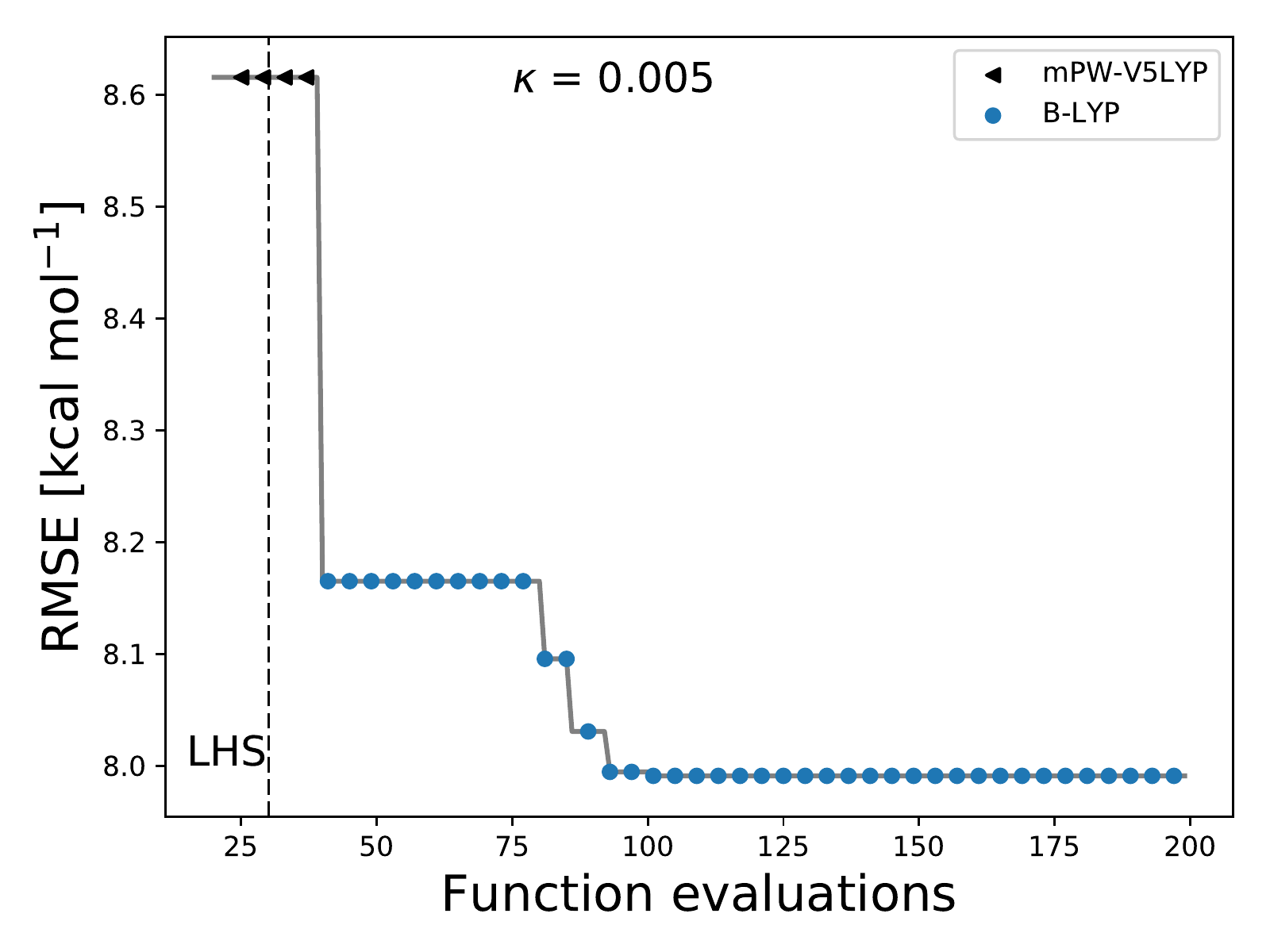}
\includegraphics[width=0.45\textwidth]{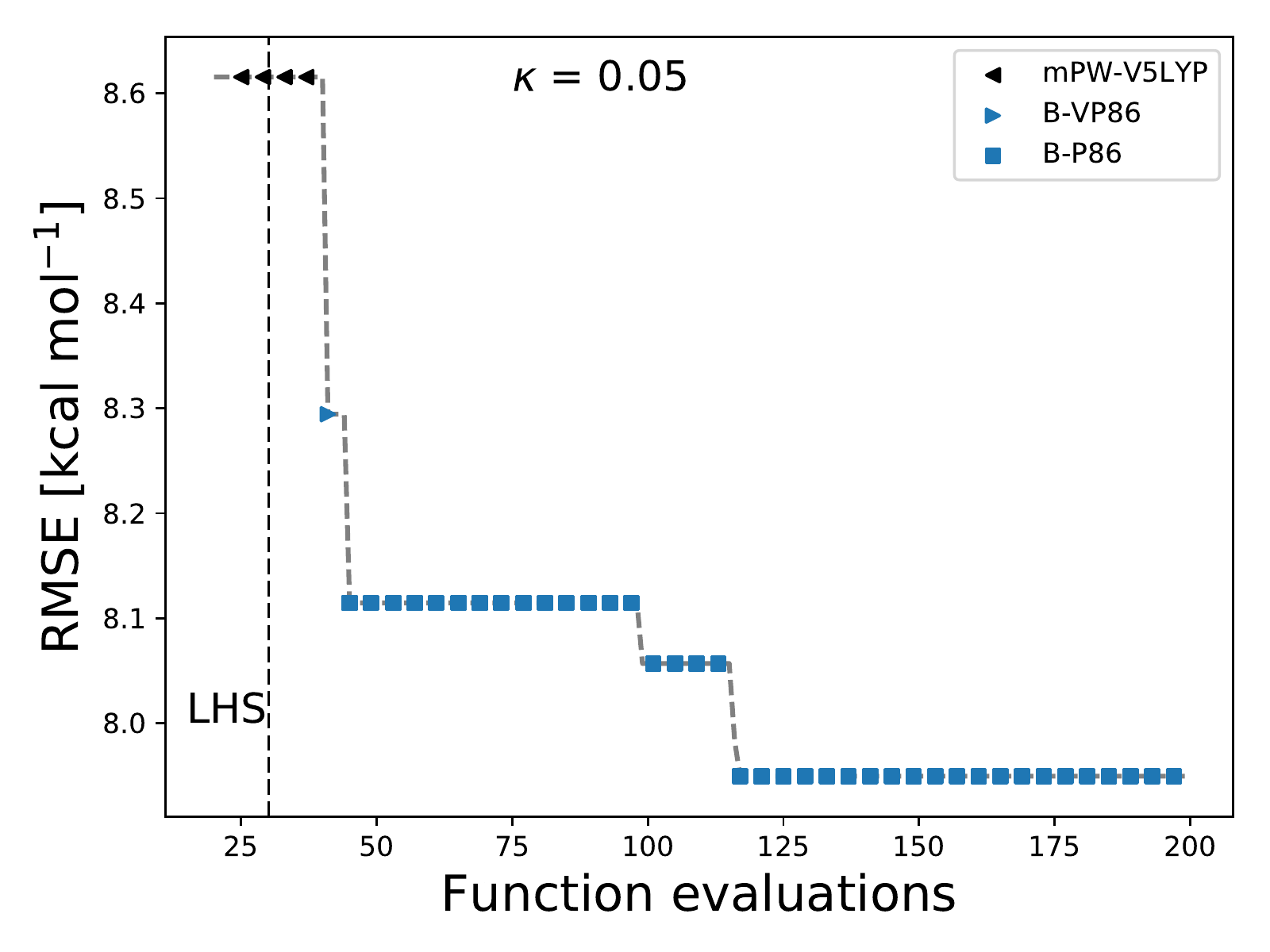}
\includegraphics[width=0.45\textwidth]{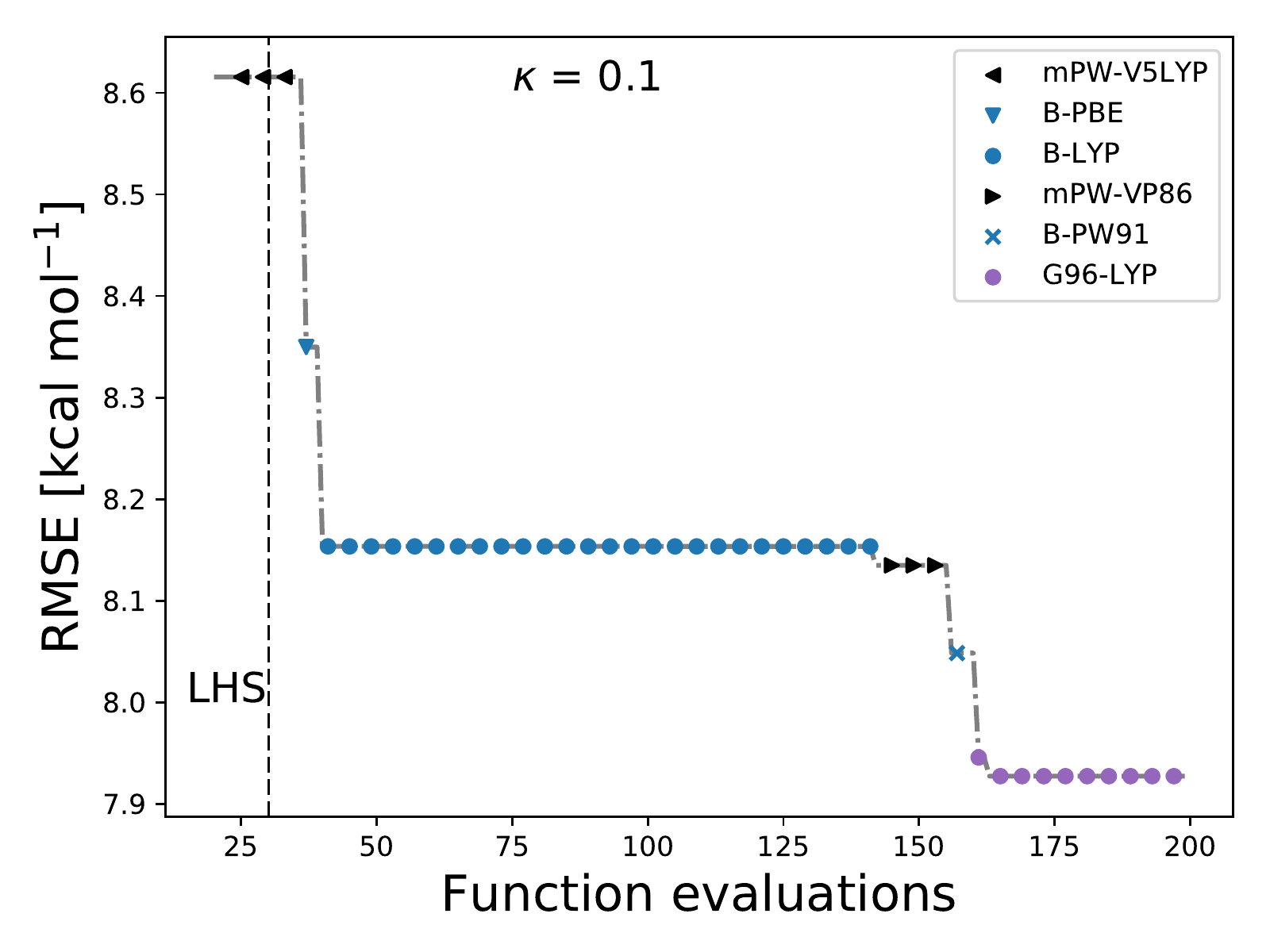}
\includegraphics[width=0.45\textwidth]{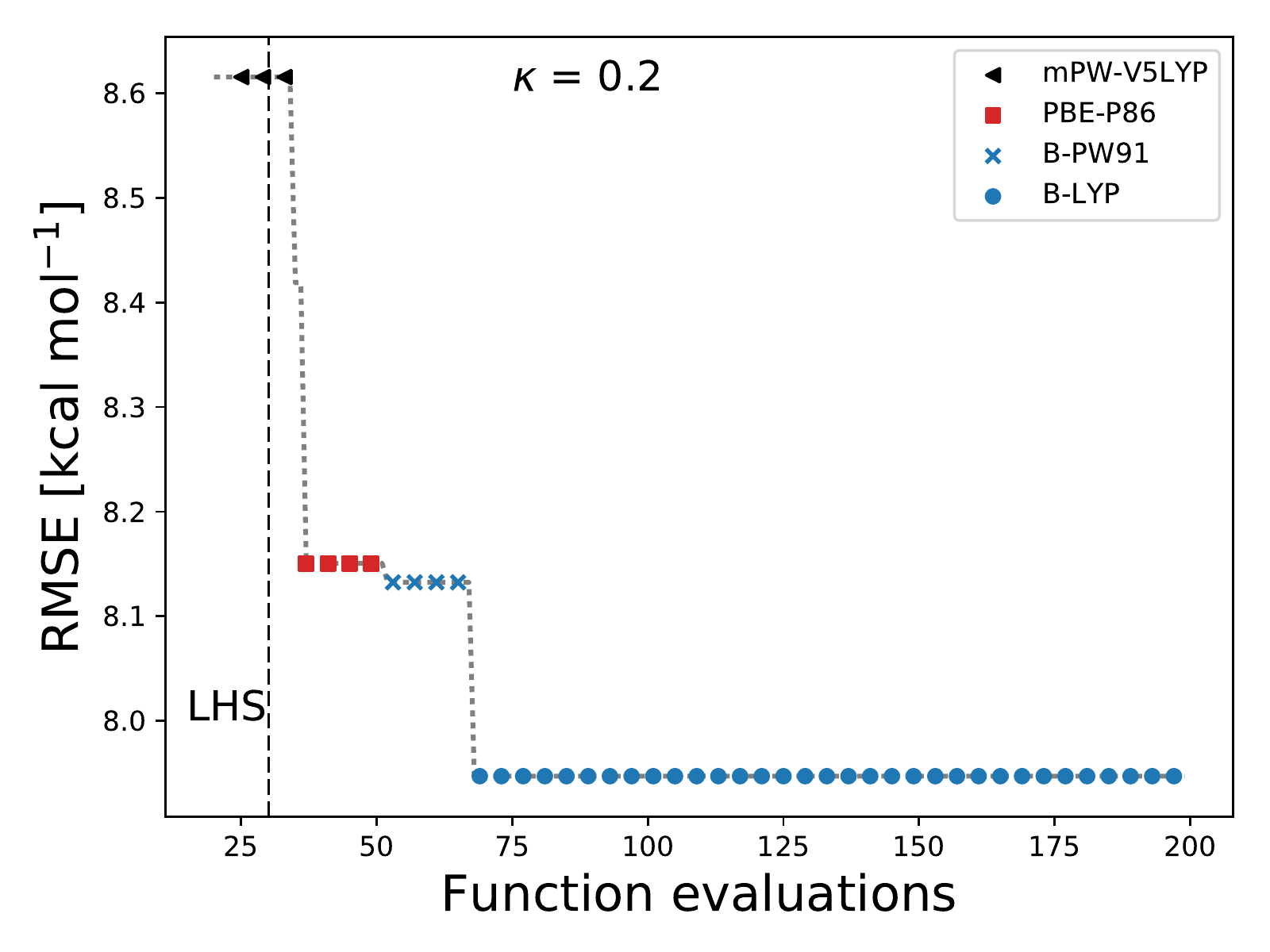}
\includegraphics[width=0.45\textwidth]{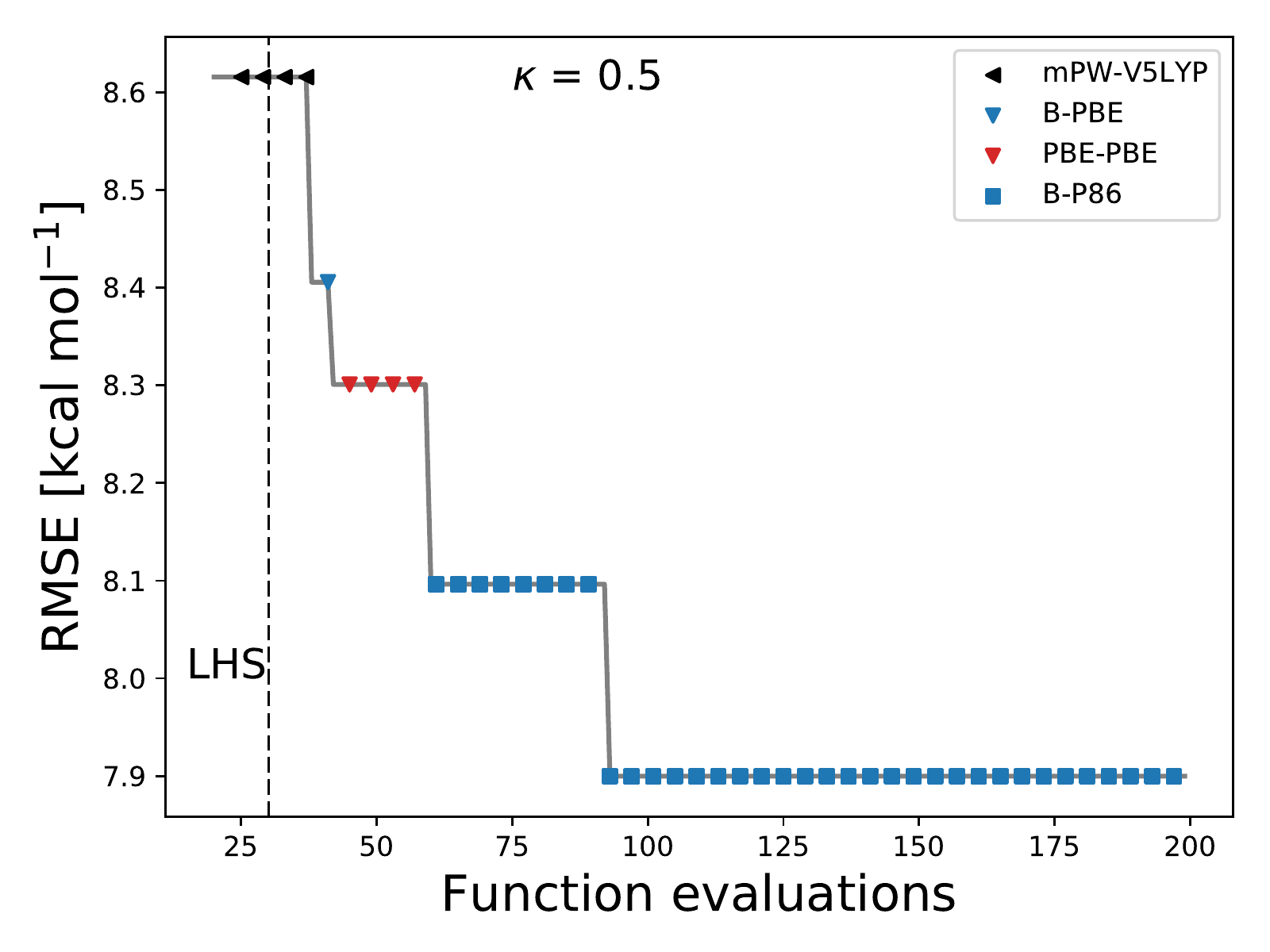}
\caption{The lowst RMSE observed value as a function of the iterations in the BO algorithm and the UCB acquisition function with different value of $\kappa$. 
The symbols represent the XC functional selected by BO with the lowest RMSE at each iteration.
The floor function was used during the optimization of the acquisition function. 
For each BO optimization we used the same 30 initial points sampled with LHS, including the XC functional. 
The RMSE is computed using Eq. \ref{eqn:loss_dft}, where $R_{m_i}$ are atomization energies of the G1 molecules \cite{G1,G1_1,G1_2}.
  }\label{fig:bo_acq_ucb_diff_k}
\end{figure}

We consider six different exchange and five different correlation functionals; $E_X = [\text{B, PW91, mPW, G96, PBE}]$ and $E_C = [\text{LYP, P86, PW91, VP86, V5LYP}]$. We label each $E_X$ and $E_C$ with different sequential integer numbers; for example for the mPW-P86 functional $E_X=2$ and $E_C = 1$.  
The points proposed in each of the BO algorithm described one $E_X$ and $E_C$ functional and the DF parameters, $[\mathbf{z},\mathbf{x}] = [[E_X,E_C],a_0,a_X,a_C]$.
During the optimization of the acquisition function we replaced the continuous values of the first two components of $\mathbf{x}$ for the with the closest integers using the floor function,
\begin{eqnarray}
[\mathbf{z},\mathbf{x}] = [\lfloor{1.45\rfloor},\lfloor{2.35\rfloor},0.15,0.8,0.01] = [1,2,0.15,0.8,0.01] = [\text{PW91},\text{PW91},0.15,0.8,0.0]. 
\end{eqnarray}
The 30 initial points were also sampled using the LHS algorithm. 
Each point is a 5-dimensional vector which first two components are the DF model, $E_X$ and $E_C$. 
We considered the impact of the floor function during the optimization of the acquisition function, Table \ref{tab:sm_exc_dft_k} and Figure \ref{fig:bo_acq_ucb_diff_k}, and different values of $\kappa$.
Form Table \ref{tab:sm_exc_dft_k} we can observe that the floor function allows BO to select DF models with lower RMSE.
Furthermore, the value of $\kappa$ in $\alpha_{UCB}$ does not have a great impact during the optimization, Figure  \ref{fig:bo_acq_ucb_diff_k}.

\begin{table}[H]
\centering
\footnotesize
\caption{The lowest RMSE found with BO with the UCB acquisition function with different values of $\kappa$. 
BO jointly optimizes the functional form,  $E_X$ and $E_C$, and the DF model's free parameters, $a_0, a_X$ and $a_C$. 
The floor function was used during the optimization of the acquisition function.
The RMSE was computed with respect to atomization energy of G1 molecules \cite{G1,G1_1,G1_2}, Table \ref{tab:ae_g1}.
For each optimization we sampled 30 initial points with the LHS algorithm, including a random XC functional.}
\begin{minipage}[b]{0.4\linewidth}
\begin{tabular}{c | c c | c c c | c } 
$\kappa$ & $E_X$ & $E_C$ & $a_0$ & $a_X$  & $a_C$ & \multirow{2}{*}{RMSE} \\
& & & & & & $[$kcal mol$^{-1}]$ \\ \hline \hline
0.005 & B &  LYP &  0.1558  &   0.7269  &   0.2588  &   7.991 \\
0.05 & B &  P86 &  0.1551  &   0.7036  &   0.0010  &   7.950 \\
0.1 & G96 &  LYP &  0.1127  &   0.7794  &   0.0010  &   7.927 \\
0.2 &B &  LYP &  0.1314  &   0.8009  &   0.0879  &   7.947 \\
0.5 &B &  P86 &  0.1024  &   0.8015  &   0.0010  &   7.900 \\
\end{tabular}
\end{minipage}
\end{table}

\bibliography{reference}